\documentclass{aa}  

\usepackage{graphicx}
\usepackage{txfonts}
\usepackage{amsmath}
\usepackage[table,xcdraw]{xcolor}
\usepackage[colorlinks=true,linkcolor=blue, filecolor=blue, urlcolor=blue, citecolor=blue]{hyperref}
\usepackage{xcolor}
\usepackage{tikz, hyperref}
\usepackage{ulem}

\usepackage{orcidlink}

\begin{document}

   \title{Polydisperse Formation of Planetesimals}

   \subtitle{The dust size distribution in clumps}
   \author{Jip Matthijsse\orcidlink{0009-0001-5298-9725}
          \inst{1}
          \and
          Hossam Aly\orcidlink{0000-0002-1342-1694}\inst{1}
          \and
          Sijme-Jan Paardekooper\orcidlink{0000-0002-8378-7608}\inst{1}}

   \institute{Planetary Exploration, Technical University Delft, Kluyverweg 1 2629 HS Delft, The Netherlands\\
            \email{j.p.matthijsse@tudelft.nl}
             }

 \date{Received XXXX; accepted YYYY}

 
  \abstract
   {}
    {To form km-sized planetesimals, the streaming instability is an efficient method for overcoming the barriers to planet formation in protoplanetary discs. The streaming instability has been extensively modelled by hydrodynamic simulations of gas and a \emph{single} dust size. However, more recent studies considering a more realistic case of a particle size distribution show that this will significantly decrease the growth rate of the instability. We follow up on these studies by evaluating the polydisperse streaming instability in the nonlinear regime to see if clumping can occur in the same manner as the monodisperse streaming instability and determine the size distribution in the densest dust structures.}
    {We employ 2D hydrodynamic simulations in an unstratified shearing box with multiple dust species representing an underlying continuous dust size spectrum using FARGO3D. We use the Gauss-Legendre quadrature in dust size space to calculate the drag force on the gas due to a continuous dust size distribution. These simulations are compared to previous analytical results of the polydisperse streaming instability in the linear phase. We then look at the saturated non-linear phase of the instability at the highest density regions and investigate the dust size distribution in the densest dust structures. }
    {When sampling the size distribution, the error in the growth rate converges significantly faster with the number of dust sizes using the Gauss-Legendre quadrature method than the usual uniform sampling method. In the non-linear regime, the maximum dust density reached in the polydisperse case is reduced compared to the monodisperse case. Larger dust particles are most abundant in the densest dust structure because they are less coupled to the gas and, therefore, can clump together more than the smaller dust grains. Contrary to expectations based solely on dust-gas coupling, our results reveal a distinct peak in the size distribution that arises from the size-dependent spatial segregation of the highest-density regions, where particles with the largest Stokes numbers are located just outside the densest areas of the combined dust species.}
    {The 2D, unstratified polydisperse streaming instability is less efficient than its monodisperse counterpart in generating dense clumps that may collapse into planetesimals, and in the densest regions, the distinct dust size distribution could relate to the size distribution that ends up in the planetesimal and can mimic the size distribution of dust growth.}

   \keywords{hydrodynamics – instabilities – methods: numerical – planets and satellites: formation – protoplanetary disc}

   \maketitle

\section{Introduction}

\par Planets form in protoplanetary discs, where solid particles with sizes of around $0.1 \ \mathrm{\mu m}$ need to clump together to form planet-sized objects of around $10,000 \ \mathrm{km}$. However, this process is not trivial, and certain barriers impede the growth of the dust. At small sizes, the growth of the particles is dominated by collisional growth,  which can lead to maximum dust sizes of $\sim \! \mathrm{cm}$, depending on local disc conditions \citep{Birnstiel2011A&A...525A..11B}. The dust also feels a drag from the gas that orbits at sub-Keplerian velocities, which causes the dust to lose angular momentum and radially drift inwards towards the host star. The drift speed is dependent on the particle size such that for meter-sized objects, the drift timescale becomes shorter than the timescale of collisional growth, making it very difficult for particles to grow larger than one meter before they drift into the host star \citep{Weidenschilling1977Ap&SS..51..153W, Nakagawa1986Icar...67..375N}.

\par Radial drift can also trigger instabilities that can aid in dust growth. The relative velocities between the dust and gas can concentrate dust particles until densities exceed the threshold for gravitational collapse, leading directly to km-sized bodies unaffected by fast radial drift. One of the most prominent drift-induced instabilities is the streaming instability \citep[SI, e.g.][]{Youdin2005ApJ...620..459Y, Youdin2007ApJ...662..613Y, Johansen2007Natur.448.1022J, Nesvorny2019NatAs...3..808N, Li2021ApJ...919..107L}. The SI has been extensively modelled using only a \emph{single} dust size (monodisperse), either with a fluid-fluid or fluid-particle approach. More recent studies considering a more realistic case of a particle size distribution \citep[first explored in][]{Bai2010ApJ...722.1437B} also show that this can dampen the instability in certain regimes \citep[see][]{Yang2021MNRAS.508.5538Y, Rucska2023MNRAS.526.1757R}. Studies of the linear phase show that the Polydisperse Streaming Instability (PSI) case will significantly decrease the growth rate of the monodisperse streaming instability (mSI) \citep{Krapp2019ApJ...878L..30K, Paardekooper2020MNRAS.499.4223P, Zhu2021MNRAS.501..467Z, McNally2021MNRAS.502.1469M}.

\par In this paper, we aim to build upon the linear analysis of the PSI by investigating the nonlinear regime and determining which dust sizes will end up in the overdensities formed by the instability during the saturated phase. This work will be a continuation of the three previous PSI papers \citep{Paardekooper2020MNRAS.499.4223P, Paardekooper2021MNRAS.502.1579P, McNally2021MNRAS.502.1469M}, focusing on the nonlinear regime using hydrodynamical simulations.

\par Other studies of the PSI in the nonlinear regime \citep{Bai2010ApJ...722.1437B, Yang2021MNRAS.508.5538Y, Schaffer2018A&A...618A..75S, Schaffer2021A&A...653A..14S, Rucska2023MNRAS.526.1757R} cover a large part of the parameter space, evaluating PSI with a fluid-particle treatment, and find dust clumps with densities exceeding the local Roche density, which may subsequently collapse into planetesimals. \cite{Yang2021MNRAS.508.5538Y} considered the nonlinear evolution of the PSI using a discrete number of dust sizes. In this study, we take the dust size distribution to be a continuous limit using a fluid-fluid treatment and only look at the high dust-to-gas ratio regime $\mu > 1$ and smaller dust sizes that are well coupled to the gas. As in \citet{Yang2021MNRAS.508.5538Y}, we will ignore the vertical component of the stellar gravity and work in an unstratified local model of the disc. This simplified model allows us to make direct contact with linear theory \citep[see][]{Krapp2019ApJ...878L..30K, Paardekooper2020MNRAS.499.4223P, Zhu2021MNRAS.501..467Z} and further our understanding of polydisperse drag instabilities. In this regime, we focus on the evolution of the dust size distribution to see if the PSI has a distinct size distribution at the highest density regions that could be related to the size distribution that ends up in planetesimals. This distribution could in turn be anchored to detailed observations of the Solar system \citep[e.g.][]{Mottola2015Sci...349b0232M, Blum2017MNRAS.469S.755B, Fulle2017MNRAS.469S..39F, Simon2018E&PSL.494...69S, Simon2022arXiv221204509S}.

\par The heuristic model behind SI in the low dust-to-gas ratio regime $\mu < 1$ is a Resonant Drag Instability \citep[RDI,][]{Squire_a_2018ApJ...856L..15S}. The gas is supported by pressure. The forces and, thus, acceleration of the gas are different from that of the dust. This leads to a drift velocity $\delta v$, which causes an acceleration acting on the dust. In the Epstein regime \citep{Epstein1924PhRv...23..710E}, this acceleration $\mathbf{\alpha}_\mathrm{drag, d}$ is linearly dependent on the gas and dust velocity difference and the dust particles stopping time $\tau_\mathrm{s}$, such that:
\begin{equation}
    \tau^\mathrm{Ep}_\mathrm{s} = \sqrt{\frac{\pi}{8}}\frac{a \rho_\mathrm{b}}{c_\mathrm{s} \rho_\mathrm{g}},
    \label{eq:stoppingtime}
\end{equation}
where $a$ is the size of the particle, $\rho_\mathrm{b}$ the bulk density of the dust particle, $\rho_\mathrm{g}$ the gas density and $c_\mathrm{s}$ the soundspeed. This leads the terminal velocity of the dust to always be in the direction of the pressure maxima. This principle can be used to create overdensities that grow exponentially if there is a perturbation in the gas, and its wave velocity $\omega_\mathcal{F}(\mathbf{k})$ matches the projected streaming velocity of the dust relative to the gas,
\begin{equation}
    \mathbf{w}_\mathrm{s} \cdot \hat{\mathbf{k}} = \omega_\mathcal{F}(\mathbf{k}),
    \label{eq:resonant}
\end{equation}
with $\mathbf{w}_\mathrm{s}$ the size-dependent velocity of dust relative to the gas and $\mathbf{k}$ the wavenumber of the perturbation. The drag feedback on the gas increases the pressure maximum, leading to exponential growth if the projected wave velocity of the pressure perturbation is the same as the dust drift \citep{Squire_b_2018MNRAS.477.5011S, Squire2020MNRAS.498.1239S, Magnan2024MNRAS.529..688M}. From the definition of the RDI, it follows that the instability must be: (1) unstable for all $\mu$, (2) have growth rates maximized when the ‘resonant condition' $\mathbf{w}_\mathrm{s} \cdot \mathbf{k} = \omega_0 = \Omega \cdot \mathbf{k}$ is satisfied, and (3)  growth rate in the linear regime scales as $\Im{(\omega)} \sim \mu^{1/2}$ rather than $\mu$. However, the drift velocity of the dust strongly depends on the dust size, making the RDI less effective for the distribution of dust sizes if there is no distinct resonant velocity \citep{Krapp2019ApJ...878L..30K, Paardekooper2020MNRAS.499.4223P, Paardekooper2021MNRAS.502.1579P, Zhu2021MNRAS.501..467Z, McNally2021MNRAS.502.1469M} \footnote{This will be further explored in Paardekooper \& Aly (in prep.), where it is shown that the RDI streaming instability is particularly negatively affected by the size distribution.}.

\par There is a second regime at high dust-to-gas ratio with higher growth rates where the PSI can grow. Importantly, the high-$\mu$ streaming instability is not an RDI and does not follow the RDI definitions from \citet{Squire_b_2018MNRAS.477.5011S}.  If the dust-to-gas ratio is higher than one ($\mu >1$). Then, there are fast-growing, high-$k$ modes that are traditionally also called the ‘streaming instability’ but are fundamentally different and share none of these RDI features and correlations that appear in low dust-to-gas ratio SI\footnote{See for example, the difference between model Af (the fast-growing instability, run with $\mu=2$ and in the high-$\mu$ SI regime and As (the slow instability with $\mu=0.2$) in \citet{Yang2021MNRAS.508.5538Y}.}. Instead, the high dust-to-gas ratio regime is more akin to a forced harmonic oscillator.

\par In this study, we aim to explore how different dust sizes clump together in the saturated regime of the instability and what the size distribution will be in the highest-density clumps. Section \ref{sec:eq_of_motions} discusses the governing equations in the local frame as well as the equilibrium state. In Section \ref{sec:numerical_aprouch}, we discuss the method of discretization of size distribution and list the numerical setups of the simulations. In Section \ref{sec:linear}, we present the growth rates of different runs and compare their growth rates to analytical work. The non-linear regime and its properties are studied in Section \ref{sec:non_linear} together with a convergence study. The simulations with different setups are used in a parameter study to investigate the effect of varying diffusion coefficient, size distribution shapes and dust-to-gas ratios in Section \ref{sec:parameterstudy}. In Section \ref{sec:substructure}, we discuss how substructures arise within the densest clumps. In Section \ref{sec:discussion}, we discuss the possible implication of the approximations made, and we summarize our findings and the implication of the found substructures in the densest clumps on planet formation.

\section{Polydisperse equations of motion}\label{sec:eq_of_motions}

\par A numerical approach is necessary to study the PSI's nonlinear evolution. We can use the Euler equations that govern the evolution of the mixture and dynamics of the gas,
\begin{align}
    \partial_t \rho_\mathrm{g} + \nabla \cdot (\rho_\mathrm{g} \mathbf{v}_\mathrm{g}) &= 0, \label{eq:euler1}\\
    \partial_t \mathbf{v}_\mathrm{g} + (\mathbf{v}_\mathrm{g} \cdot \nabla) \, \mathbf{v}_\mathrm{g} &= - \frac{\nabla p}{\rho_\mathrm{g}} + \boldsymbol{\alpha}_\mathrm{g} + \boldsymbol{\alpha}_{\mathrm{drag}, \, \mathrm{g}}, \label{eq:euler2}
\end{align}
where $p$ is the gas pressure, $\boldsymbol{\alpha}_g$ are the forces acting on the gas except the drag force of the dust denoted with $\boldsymbol{\alpha}_{\mathrm{drag}, \,g}$. 

\par We can take the velocity moments of the Boltzmann equation to derive the fluid equations that describe the evolution of the dust particles:
\begin{align}
    \partial_t \sigma + \nabla \cdot (\sigma \mathbf{u}) &= 0, \label{eq:dustfluid1}\\
    \partial_t  \mathbf{u} + (\mathbf{u} \cdot \nabla) \, \mathbf{u} &= \boldsymbol{\alpha}_\mathrm{d} (\mathbf{u}) + \boldsymbol{\alpha}_{\mathrm{drag}, \, \mathrm{d}} (\mathbf{u}), \label{eq:dustfluid2}
\end{align}
here, $\mathbf{u}$ is defined as the size-dependent bulk velocity ($\mathbf{u} \equiv \langle \mathbf{v}_\mathrm{d} \rangle_\mathbf{v}$), $\sigma$ the size density, and $\boldsymbol{\alpha}_\mathrm{d}(\mathbf{u})$ the acceleration due to external forces and $\boldsymbol{\alpha}_{\mathrm{drag}, \, \mathrm{d}} (\mathbf{u})$ from the drag between the dust and the gas  \citep[see,][for full derivation]{Paardekooper2021MNRAS.502.1579P}. In this derivation, we neglect the divergence of the stress tensor. This can only be done if we are in the regime of the \emph{fluid approximation}. The fluid approximation is only valid for particles for which the coupling to the gas is strong enough, requiring Stokes number $\tau_\mathrm{s} \ll 1$ \citep{Garaud2004ApJ...603..292G, Jacquet2011MNRAS.415.3591J}. Compared to the mSI, the backreaction on the gas is dependent on the \emph{size density} $\sigma$, given by,
\begin{equation}
    \sigma(\mathbf{x}, t, a) = \rho_\mathrm{b} V(a) \int f(\mathbf{x}, \mathbf{v}_\mathrm{d}, a, t) \, \mathrm{d}\mathbf{v},
    \label{eq:size_density}
\end{equation}
with $V(a)$ the volume of the dust particle ($V(a) = 4\pi a^3/3$ for spherical dust particles). The size density is the mass density between $a$ and $a + \mathrm{d}a$, such that, 
\begin{align}
    \rho_\mathrm{d}  &= \int \sigma \mathrm{d}a, \label{eq:dust_dens}\\
    \rho_\mathrm{d} \mathbf{v}_\mathrm{d} &= \int \sigma \mathbf{u} \mathrm{d}a,\label{eq:dust_momentum}
\end{align} 
with $\mathbf{v}_\mathrm{d}$ the bulk velocity of the dust. We can use these equations together with the definition of the acceleration of the dust  $\mathbf{\alpha}_\mathrm{drag, d}$, to derive an expression for the total momentum transfer between the dust and gas,
\begin{equation}
    \rho_\mathrm{g} \boldsymbol{\alpha}_\mathrm{drag, g} =  \int \sigma(a) \frac{\mathbf{u}- \mathbf{v}_\mathrm{g}}{\tau_\mathrm{s}(a)}\mathrm{d}a.
    \label{eq:backreaction}
\end{equation}

\subsection{Local frame}

\par We simulate the PSI within a shearing box \citep{Goldreich1965MNRAS.130..125G}, a co-rotating frame at an orbiting radius $r_0$ with Cartesian coordinates ($\hat{\mathbf{r}} \rightarrow \hat{\mathbf{x}}; \, \hat{\mathbf{\phi}} \rightarrow \hat{\mathbf{y}}; \, \hat{\mathbf{z}} \rightarrow \hat{\mathbf{z}}$). The transformation from the inertial frame will introduce inertial forces within the shearing box, the Coriolis and tidal forces. These forces are included through an effective potential $\Phi = -S\Omega x^2$, with $S$ the shearing rate, in a Keplerian disc $S = \frac{3}{2}\Omega$ with $\Omega$ the angular velocity. A global pressure gradient ($\partial P/\partial r$) exists throughout the disc in the local frame. This pressure gradient is described with the parameter\footnotemark;
\begin{align}
    \eta = \frac{1}{2 \rho_\mathrm{g}} \frac{\partial P}{\partial r} \sim \frac{c_\mathrm{s}^2}{r_0}.
    \label{eq:pressure_gradient}
\end{align}
\footnotetext{Note that we use a dimensional pressure support parameter $\eta$ similar to \citet{Paardekooper2020MNRAS.499.4223P, Paardekooper2021MNRAS.502.1579P, McNally2021MNRAS.502.1469M}. This parameter is related to the non-dimensional parameter defined in \citet{Youdin2005ApJ...620..459Y} by $\eta = r_0 \Omega^2 \eta_\mathrm{YG}$. This is a largely cosmetic choice allowing us to use a length scale of $\eta/\Omega^2$, without using $r_0$ the orbital radius of the shearing box.}
This means that the background acceleration on the gas, without the correction of the dust, can be given by 
\begin{align}
    \boldsymbol{\alpha}_\mathrm{g} = 2 \eta \hat{\mathbf{x}} - 2 \mathbf{\Omega} \times \mathbf{v}_\mathrm{g} - \nabla \Phi.
    \label{eq:background_accelaration}
\end{align}

\par  Taking into account the equations from the frame transformation \eqref{eq:background_accelaration}, momentum transfer \eqref{eq:backreaction} and the original equation governing the motions and mixture of gas and dust \eqref{eq:euler1}, \eqref{eq:euler2}, \eqref{eq:dustfluid1} and \eqref{eq:dustfluid2}, will give the equations of motion within a shearing-box:
\begin{align}
    \partial_t \rho_\mathrm{g} + \nabla \cdot (\rho_\mathrm{g} \mathbf{v}_\mathrm{g}) &= 0, \label{eq:shearingbox_gas1}\\
    \partial_t \mathbf{v}_\mathrm{g} + (\mathbf{v}_\mathrm{g} \cdot \nabla) \, \mathbf{v}_\mathrm{g} &= 2\eta \hat{\mathbf{x}} - \frac{\nabla p}{\rho_\mathrm{g}} -2\Omega\times \mathbf{v}_\mathrm{g} - \nabla \Phi \notag\\
    & \quad + \frac{1}{\rho_\mathrm{g}} \int \sigma(a) \frac{\mathbf{u}-\mathbf{v}_\mathrm{g}}{\tau_\mathrm{s}(a)}\mathrm{d}a,
    \label{eq:shearingbox_gas2}\\
    \partial_t \sigma + \nabla \cdot (\sigma \mathbf{u}) &= 0, \label{eq:shearingbox_dust1}\\
    \partial_t \mathbf{u} + (\mathbf{u} \cdot \nabla) \, \mathbf{u} &= -2\Omega \times \mathbf{u} - \nabla \Phi - \frac{\mathbf{u}-\mathbf{v}_\mathrm{g}}{\tau_\mathrm{s}(a)}
    \label{eq:shearingbox_dust2}
\end{align}

 \subsection{Equilibrium state}

 \par From the shearing box equations \eqref{eq:shearingbox_gas1}-\eqref{eq:shearingbox_dust2}, we can find an equilibrium solution, this will indicate the radial drift and shear velocity of the gas and dust particles. We assume an isothermal equation of state, and all quantities are constant in space (apart from the shear) and have no vertical gravity (this is thought to represent the midplane of a protoplanetary disc). Then, the equilibrium solution is given by,
 \begin{align}
     v_{\mathrm{g}, \, x} &= \frac{2\eta}{\kappa} \frac{\mathcal{J}_1}{(1+\mathcal{J}_0)^2 + \mathcal{J}_1^2},
     \label{eq:gasx_equilibrium_solution}\\
     v_{\mathrm{g}, \, y} &= -Sx- \frac{\eta}{\Omega} \frac{1+ \mathcal{J}_0}{(1+\mathcal{J}_0)^2 + \mathcal{J}_1^2},
     \label{eq:gasy_equilibrium_solution}\\
     u_x &= \frac{2 \eta}{\kappa} \frac{\mathcal{J}_1 - \kappa \tau_\mathrm{s}(a) (1+ \mathcal{J}_0)}{(1+\kappa^2 \tau_\mathrm{s}(a)^2)((1+\mathcal{J}_0)^2+\mathcal{J}_1^2)},
     \label{eq:dustx_equilibrium_solution}\\
     u_y &= -Sx -\frac{\eta}{\Omega} \frac{1 + \mathcal{J}_0 + \kappa \tau_\mathrm{s}(a)\mathcal{J}_1}{(1+\kappa^2\tau_\mathrm{s}(a)^2)((1+\mathcal{J}_0)^2+\mathcal{J}_1^2)},
     \label{eq:dusty_equilibrium_solution}\\
     v_{\mathrm{g}, \, z} &= u_z = 0,
     \label{eq:z_equilibrim_solution}
 \end{align}
 where $\kappa$ is the epicyclic frequency, and $\mathcal{J}_m$ a series of integrals given by,
 \begin{align}
     \mathcal{J}_m = \frac{1}{\rho_\mathrm{g}} \int \frac{\sigma(\kappa \tau_\mathrm{s}(a))^m}{1+\kappa^2\tau_\mathrm{s}(a)^2}\mathrm{d}a,
     \label{eq:interal_equilibrium_solution}
 \end{align}
(from \citet{Paardekooper2020MNRAS.499.4223P}; first derived in \citealt{Tanaka2005ApJ...625..414T}). These equations form the background solution in the shearing box.

\subsection{Linear Analysis}

To evaluate the growth of the PSI, we need to have a perturbation on top of the equilibrium state that grows in the form of
\begin{align}
    X(\mathbf{x}, t, a) = X^0(\mathbf{x}, a) + X^1(\mathbf{x}, t, a),
    \label{eq:state}
\end{align}
if we take the perturbation to be in the form of 
\begin{align}
    X^1(\mathbf{x}, t, a) = \hat{X}(a) \exp{\left( i \mathbf{k} \cdot \mathbf{x} -i \omega t\right)},
    \label{eq:pertubation}
\end{align} 
here $\mathbf{k} = \left( k_x, k_y, k_z\right)^T$ is the wavevector and $\omega$ is the frequency. The growth rate of the perturbation is dependent on the imaginary part of the frequency, and the instability will grow if $\Im(\omega) > 0$. We can analytically determine the value of $\omega$ by solving the equations of motion \eqref{eq:shearingbox_gas1}-\eqref{eq:shearingbox_dust2} for \eqref{eq:state}. If we want to use these equations to solve for $\omega$, this will constitute an integral eigenvalue problem for eigenvalue $\omega$. This eigenvalue problem can be solved by either a discrete solver \citep{Krapp2019ApJ...878L..30K} or by using a root finding method; both of these methods are described in \citet{Paardekooper2021MNRAS.502.1579P} and an implementation in the publicly available Python package \texttt{psitools}\footnote{\url{https://doi.org/10.5281/zenodo.4663587}} \citep{Paardekooper2020MNRAS.499.4223P, Paardekooper2021MNRAS.502.1579P, McNally2021MNRAS.502.1469M}.

\par We want to further simplify the problem by only using dimensionless units by choosing a time scale $\Omega^{-1}$ and a length scale $\eta /\Omega^2$. When the time scale is $\Omega^{-1}$  the Stokes number is the non-dimensional stopping time $\mathrm{St} \equiv \Omega \tau_\mathrm{s}$, the parameters governing the system are now: the non-dimensional wave vector $\mathbf{K} = \mathbf{k} \eta /\Omega^2$, the non-dimensional sound speed $c_\mathrm{s}/(\Omega \eta)$, the shear parameter $S/\Omega$, the dust-to-gas ratio $\mu= \rho_\mathrm{d}^0/\rho_\mathrm{g}^0$ and the size density $\sigma^0(a)$. We use the relation between Stokes number and dust size \eqref{eq:stoppingtime} to express the size density in the non-dimensional stopping time $\sigma^0(\tau_\mathrm{s})$.

\section{Numerical approach}\label{sec:numerical_aprouch}

\par We evaluate the nonlinear regime with numerical simulation using the hydrodynamical code \texttt{FARGO3D} \citep{Benitez-Llambay2016ApJS..223...11B, Benitez-Llambay2019ApJS..241...25B} to evaluate the clumping of different dust sizes of the PSI.

\subsection{Approximating the drag integral}\label{sec:gaussianlegendrequadrature}

\texttt{FARGO3D} allows for multiple dust fluids in a single simulation. In this multifluid picture, where a discrete number of dust sizes is considered, the backreaction on the gas is dictated by momentum conservation \citep{Benitez-Llambay2019ApJS..241...25B}:
\begin{align}
  \rho_{\rm g}{\bf a}_{\rm drag,g} = \sum_j \frac{\rho_{{\rm d},j}}{\tau_{{\rm s},j}}\left({\bf u}_j-{\bf v}_{\rm g}\right),
  \label{eq:drag_discrete}
\end{align}
where the sum is over all dust species, and subscripts $j$ indicate quantities belonging to the $j$th dust species. In the polydisperse picture, where instead of a discrete number of sizes, we have a continuous size distribution, we can view the above equation as an approximation to the integral (\ref{eq:backreaction}). This can be made explicit by defining $\rho_{{\rm d},j}=\sigma(a_j)\Delta a$, in the case of constant size bins of width $\Delta a$. The size distribution will usually span many orders of magnitude, in which case it is advantageous to choose size bins that have a constant size in log space, but the fundamental principle remains that the integral (\ref{eq:backreaction}) is approximated by a (middle) Riemann sum or, in other words, the midpoint rule. This method is commonly used to approximate the drag integral \citep[see e.g.][]{Bai2010ApJ...722.1437B, Krapp2019ApJ...878L..30K, Schaffer2018A&A...618A..75S, Zhu2021MNRAS.501..467Z, Yang2021MNRAS.508.5538Y} and we will henceforward refer to this method as the discrete method.

If we do not start with a multifluid approach, but assume from the start that we have a continuous size distribution, we need an accurate but robust way to calculate the drag integral (\ref{eq:backreaction}). Without a-priori knowledge about the integrand, Gauss-Legendre quadrature (GL) is a promising step forward from the midpoint rule. It is still convenient to work in log-size space, but rather than choosing bins of constant widths, the integration nodes are now given by the roots of the $n$th Legendre polynomial if we consider $n$ integration nodes. If we, for simplicity, integrate over $a$ rather than in log space, we have that
\begin{align}
  \int_{a_{\rm min}}^{a_{\rm max}} \sigma(a) \frac{{\bf u}(a)-{\bf v}_{\rm g}}{\tau_{\rm s}}{\rm d} a \approx \sum_{j=0}^n w_j \sigma(a_j) \frac{{\bf u}(a_j)-{\bf v}_{\rm g}}{\tau_{{\rm s},j}},
\label{eq:drag_gauss}
\end{align}
where $w_j$ are the GL weights and $a_j$ are the integration nodes, both appropriately adjusted for the integration range. If we compare (\ref{eq:drag_gauss}) to (\ref{eq:drag_discrete}), we see that if we \emph{define} the dust densities as the product of the size density at the integration node and the corresponding weight, the equations are identical. This means that the GL integration method can be implemented in all hydrodynamics codes using pressureless fluids by choosing the appropriate sizes and redefining dust densities. Importantly, no changes in the internal workings of the code are necessary.

We can define the appropriate size and redefined dust densities with \texttt{scipy.special.roots\_legendre} \citep{Virtanen2020NatMe..17..261V}. This function gives the roots of the Legendre polynomial $x_j$ and weights $w_j$ between $[-1;1]$. If we want to work in log-size space we can map these notes to the correct Stokes number using,
\begin{align}
    \tau_{\mathrm{s}, \, j} = \tau_\mathrm{s, \, min} \left( \frac{\tau_\mathrm{s, \, max}}{\tau_\mathrm{s, \, min}} \right)^{(x_j+1)/2}
    \label{eq:gausslegendre_tau}
\end{align}
and the redefined dust density by,
\begin{align}
    \rho_{\mathrm{d}, \, j} = \frac{1}{2} \ln{\left( \frac{\tau_\mathrm{s, \, max}}{\tau_\mathrm{s, \, min}} \right)} \ w_j \ \tau_{\mathrm{s}, \, j} \ \sigma(\tau_{\mathrm{s}, \, j}).
    \label{eq:gausslegendre_rho}
\end{align}

\begin{figure}
    \centering
    \includegraphics[width=1\linewidth]{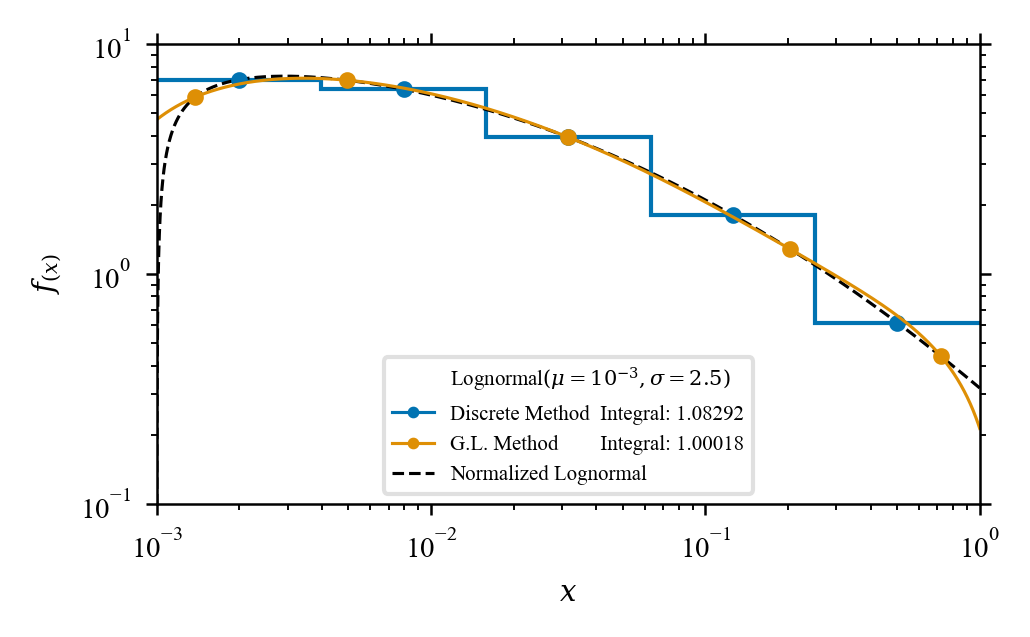}
    \caption{Integration of a Normalized Lognormal distribution (dashed line) using a discrete method (blue line) and a Gauss-Legendre quadrature (orange line).}
    \label{fig:example_GL}
\end{figure}

We give an example of the differences between the integration methods in Figure \ref{fig:example_GL} by integrating a lognormal distribution using five points with the discrete and the GL method. The integration error using the discrete method is $8.292\%$ and $0.018\%$ using the GL method. This error will be smaller when using more integration points, but GL is more accurate using fewer integration points, which is less computationally expensive.

It is worth noting that, while GL and the discrete method have the same form, they differ fundamentally. When using GL, we let go of the concept of size bins: all we have is integration nodes. While the sum over all dust 'densities' $\rho_{{\rm d},j}=w_j \sigma(a_j)$ gives the total dust density, as in the discrete case, this does not hold for partial sums. In other words, the GL method is optimized for integrating over the full-size distribution.
 
\subsection{The initial size distribution}\label{sec:sizedistribution}

 \par The size distribution $\sigma(a)$ in a protoplanetary disc can be expected to be inherited from the interstellar medium. The particle size distribution of the interstellar medium follows a power law, given by MRN distribution \citep{Mathis1977ApJ...217..425M, Draine1984ApJ...285...89D}
 \begin{equation}
    \sigma_\mathrm{MRN}(a, \beta) \propto a^{3+\beta}, \quad \beta = -3.5 .
     \label{eq:MRN}
 \end{equation}
 A power law will also follow from collisional evolution if we expect \emph{pure} fragmentation or \emph{pure} coagulation, making it a reasonable approximation for the size distribution when studying the PSI \citep[e.g. used in][]{Krapp2019ApJ...878L..30K, Paardekooper2020MNRAS.499.4223P, McNally2021MNRAS.502.1469M, Yang2021MNRAS.508.5538Y}. When using dimensionless units to set up the simulations, the size distribution will be expressed in non-dimensional stopping time (Stokes number) $\tau_\mathrm{s}$, the stopping times can be converted back to dust sizes using \eqref{eq:stoppingtime} and a disc model, e.g. as a rough guide, a meter-sized boulder orbiting in the Minimum Mass Solar Nebula (MMSN) at one Astronomical Unit (AU) has a Stokes number of order unity \citep{Weidenschilling1977Ap&SS..51..153W}.

\par In our numerical study of the PSI, we want to explore a similar parameter space as was covered in the analytical analysis of the PSI in the linear regime by \citet{McNally2021MNRAS.502.1469M}. For our standard case, we stay in the high-$\mu$ regime and take the dust-to-gas ratio $\mu = 3$, although we vary this later (see Section \ref{sec:dust_gas_ratio}). 

\par The largest Stokes number $\tau_\mathrm{s, \, max}$ in the distribution needs to be small enough that the fluid approximation holds. This approximation breaks down for $\tau_\mathrm{s} \gtrsim 1 \, \Omega^{-1}$, therefore we take $\tau_\mathrm{s, \, max} = 10^{-1} \, \Omega^{-1}$. Although the dust size distribution in a protoplanetary disc can cover orders of magnitude, the error in the back-reaction of the dust is dependent on the sample rate of the distribution and is dominated by the larger dust sizes. Therefore we sample $\mathcal{O}(2)$ with $\tau_\mathrm{s} \in [10^{-3}; 10^{-1}] \ \Omega^{-1}$ \citep[similar to the Stokes range in run Af and As in][]{Yang2021MNRAS.508.5538Y}.

\subsection{Setup}

We will simulate the PSI with a shearing box in two spatial dimensions $(\hat{\mathbf{x}} \text{ \& } \hat{\mathbf{z}})$, where the vertical stellar gravity is neglected (unstratified and isothermal). This means that the boundaries of the shearing box can be periodic. The background velocities of the gas and dust are given by the equilibrium solution \eqref{eq:gasx_equilibrium_solution}-\eqref{eq:interal_equilibrium_solution}. The gas and dust densities and velocities are perturbed by a small wave given by,
\begin{align}
    X^1(\mathbf{x},\mathbf{K}, \alpha) &= \hat{X} \cdot (\Re{(\delta \hat{f}(K_x, K_z, \alpha))} \cos(K_x x + K_z z) \notag \\ & \quad - \Im{(\delta \hat{f}(K_x, K_z, \alpha))}\sin(K_x x + K_z z) ).
    \label{eq:pertubation2}
\end{align}
The amplitude $\hat{X} = 10^{-5}$ for all simulations\footnote{Except the simulations of varying dust-to-gas ratio, these are perturbed with white noise in the gas, with a standard deviation of $10^{-4} c_\mathrm{s}$}, the non-dimensional wavenumber $\mathbf{K} = \left(30,0,30\right)^T$ (from Figure 9 in \citealt{McNally2021MNRAS.502.1469M} we know that at this wavenumber the instability grows $\Im{(\omega)} > 0$, even if there is viscosity $\alpha$) and  $\delta \hat{f}(\mathbf{K}, \alpha, \mu, \sigma)$ are the $4(N+1)$-dimensional (complex) eigenvectors. These eigenvectors are associated with the (complex) eigenvalue $\omega(\mathbf{K}, \alpha, \sigma)$ and the (complex) mode frequency $\omega$ is calculated for a specific wave number $\mathbf{K}$ and viscosity parameter $\alpha$ with \texttt{psitools.psimode}. This setup is the same as the test problem LinA from \citet{Youdin2007ApJ...662..613Y}.

\par The size of the shearing box is given in terms of the non-dimensional wavenumber $L_{x,z} = 2\pi/K_{x,z}$. In this study, we perturb the system with one wavenumber and take the size of the box to be the length of the wave ($L_{x,z}= 2\pi/30 \ \mathrm{\eta/\Omega^2}$). The spatial resolution of the hydro-code is an important parameter of convergence; the linear regime of the instability only needs modest spatial resolution, e.g. only needing $4\times 4$ for mSI. However, in a nonlinear regime, the spatial resolution governs the limit of the smallest resolved clumping and maximum density. The spatial resolution is, therefore, always a compromise between accuracy and computation time. In this study, we do a larger parameter sweep at a resolution of $256 \times 256$ and a convergence study up to a resolution of $1024 \times 1024$. The specific combination of spatial resolution $N_\mathrm{grid}$, number of dust species $n_\mathrm{d}$ and other parameters for different simulations are defined in Table \ref{tab:simulations}.

\begin{table}
    \caption{List simulation runs}     
    \label{tab:simulations}      
    \centering 
    \setlength{\tabcolsep}{2pt}
    \renewcommand{\arraystretch}{1.2}
    \begin{tabular}{l c c c c c c c }     
        \hline\hline       
        \textbf{Run Name} & $\mathbf{n_\mathrm{d}}$ & $\mathbf{N}_\mathrm{grid}$ & \textbf{Sample} & $\boldsymbol{\alpha}$ & $\boldsymbol{\tau}_\mathrm{s, max}$ & $\boldsymbol{\beta}$ & $\boldsymbol{\mu}$ \\
        \hline
        \texttt{PSI20$_\texttt{1024}$} & $20$ & $1024$ & G.L. & 0 & $10^{-1}$ & $-3.5$ & $3$ \\
        \texttt{PSI10}$_\texttt{1024}$ & $10$ & $1024$ & G.L. & $0$ & $10^{-1}$ & $-3.5$ & $3$\\
        \texttt{PSI5}$_\texttt{1024}$ & $5$ & $1024$ & G.L. & $0$ & $10^{-1}$ & $-3.5$ & $3$\\
        \texttt{mSI}$_\texttt{1024}$ & $1$ & $1024$ & G.L. & $0$ & $10^{-1}$ & $-3.5$ & $3$\\
        \hline
        \texttt{PSI10}$_\texttt{512}$ & $10$ & $512$ & G.L. & $0$ & $10^{-1}$ & $-3.5$ & $3$ \\
        \texttt{mSI}$_\texttt{512}$ & $1$ & $512$ & G.L. & $0$ & $10^{-1}$ & $-3.5$ & $3$ \\
        \texttt{PSI20} & $20$ & $256$ & G.L. & $0$ & $10^{-1}$ & $-3.5$ & $3$  \\
        \texttt{PSI10} & $10$ & $256$ & G.L. & $0$ & $10^{-1}$ & $-3.5$ & $3$ \\
        \texttt{mSI} & $1$  & $256$ & G.L. & $0$ & $10^{-1}$ & $-3.5$ & $3$\\
        \hline
        \texttt{PSI40}$_\texttt{disc.}$  & $40$ &  $256$&  Discrete & $0$ & $10^{-1}$ & $-3.5$ & $3$\\
        \texttt{PSI20}$_\texttt{disc.}$  & $20$ &  $256$&  Discrete & $0$ & $10^{-1}$ & $-3.5$ & $3$\\
        \texttt{PSI10}$_\texttt{disc.}$  & $10$ &  $256$&  Discrete & $0$ & $10^{-1}$ & $-3.5$ & $3$ \\
        \hline
        \texttt{PSI}$_{\alpha,1\mathrm{e}-8}$ & $10$ & $256$ & G.L. & $10^{-8}$ & $10^{-1}$ & $-3.5$ & $3$ \\
        \texttt{mSI}$_{\alpha,1\mathrm{e}-8}$ & $1$ & $256$ & G.L. & $10^{-8}$  & $10^{-1}$& $-3.5$ & $3$ \\
        \texttt{PSI}$_{\alpha,1\mathrm{e}-7}$ & $10$ & $256$ & G.L. & $10^{-7}$  & $10^{-1}$& $-3.5$ & $3$ \\
        \texttt{mSI}$_{\alpha,1\mathrm{e}-7}$ & $1$  & $256$ & G.L. & $10^{-7}$ & $10^{-1}$ & $-3.5$ & $3$\\
        \texttt{PSI}$_{\alpha,1\mathrm{e}-6}$ & $10$ & $256$ & G.L. & $10^{-6}$ & $10^{-1}$ & $-3.5$ & $3$\\
        \texttt{mSI}$_{\alpha,1\mathrm{e}-6}$ & $1$ & $256$ & G.L. & $10^{-6}$ & $10^{-1}$ & $-3.5$ & $3$\\
        \hline
        \texttt{PSI}$_{\tau_\mathrm{s}, 5\mathrm{e}-2}$ & $10$ & $256$ & G.L. & $0$ & $5\cdot10^{-2}$ & $-3.5$ & $3$\\
        \texttt{mSI}$_{\tau_\mathrm{s}, 5\mathrm{e}-2}$ & $1$  & $256$ & G.L. & $0$ & $5\cdot10^{-2}$ & $-3.5$ & $3$\\
        \texttt{PSI}$_{\tau_\mathrm{s}, 2\mathrm{e}-1}$ & $10$ & $256$ & G.L. & $0$ & $2\cdot10^{-1}$ & $-3.5$ & $3$\\
        \texttt{mSI}$_{\tau_\mathrm{s}, 2\mathrm{e}-1}$ & $1$  & $256$ & G.L. & $0$ & $2\cdot10^{-1}$ & $-3.5$ & $3$\\
        \hline
        \texttt{PSI}$_{\beta, -3.8}$ & $10$ & $256$ & G.L. & $0$ & $10^{-1}$ & $-3.8$ & $3$\\
        \texttt{PSI}$_{\beta, -3.2}$ & $10$ & $256$ & G.L. & $0$ & $10^{-1}$ & $-3.2$ & $3$ \\
        \hline
        \texttt{PSI}$^*_{\mu, 0.5}$ & $10$ & $256$ & G.L. & $0$ & $10^{-1}$ & $-3.5$ & $0.5$\\
        \texttt{mSI}$^*_{\mu, 0.5}$ & $1$ & $256$ & G.L. & $0$ & $10^{-1}$ & $-3.5$ & $0.5$\\
        \texttt{PSI}$^*_{\mu, 1}$ & $10$ & $256$ & G.L. & $0$ & $10^{-1}$ & $-3.5$ & $1$\\
        \texttt{mSI}$^*_{\mu, 1}$ & $1$ & $256$ & G.L. & $0$ & $10^{-1}$ & $-3.5$ & $1$\\
        \texttt{PSI}$^*_{\mu, 3}$ & $10$ & $256$ & G.L. & $0$ & $10^{-1}$ & $-3.5$ & $3$ \\
        \texttt{mSI}$^*_{\mu, 3}$ & $1$  & $256$ & G.L. & $0$ & $10^{-1}$ & $-3.5$ & $3$\\
        \hline\hline
        \multicolumn{2}{c}{$\left(\mathbf{L_x \times L_z}\right)$} & $\mathbf{c_\mathrm{s}}/(\Omega \eta)$ & $\boldsymbol{\rho}^0_\mathrm{g} $& $\boldsymbol{\tau}_\mathrm{s, min}$ & \multicolumn{2}{c}{$\mathbf{K}^*$}  & $\mathbf{\hat{X}}$ \\
        \multicolumn{2}{c}{$ \left( \frac{2\pi}{30} \times \frac{2\pi}{30} \right) \eta/\Omega^2$} &$20$ & $1$ & $10^{-3}$ & \multicolumn{2}{c}{$\left(30,0,30\right)^T$}  & $10^{-5}$\\
        \hline
    \end{tabular}
    \tablefoot{
    Simulation parameters, with $n_\mathrm{d}$ the number of dust species, $N_\mathrm{grid}$ the spatial resolution, \emph{Sample} the sampling method of the size distribution (G.L. stands for GL method and \emph{Discrete} for the discrete method), $\alpha$ the dust diffusion, $\tau_\mathrm{s, \, min}$ and $\tau_\mathrm{s, \, max}$ the domain of stopping times, $\beta$ the slope of the size distribution ($\sigma_\mathrm{MRN}(\tau_\mathrm{s}, \beta)$, see \eqref{eq:MRN}), $\mu$ the dust-to-gas ratio, $L$ the size of the shearing box, $c_\mathrm{s}$ the sound speed, $\rho^0_g$ the gas density, $\mathbf{K}$ the non-dimensional wavenumber and $\hat{X}$ the amplitude of the perturbation. All the simulations are perturbed by the same sinusoid except the simulations of varying dust-to-gas ratios that are perturbed by white noise with a standard deviation of $10^{-4} c_\mathrm{s}$.
    }
\end{table}

\section{Linear regime}\label{sec:linear}

\begin{table}
    \caption{Growth rates}
    \label{tab:growthrate}      
    \centering  
    \setlength{\tabcolsep}{3pt}
    \begin{tabular}{l l l l}   
        \hline\hline         
        Run &  $\Im{(\omega)}_\mathrm{A.}$ & $\Im{(\omega)}_\mathrm{N.}$ & $\Delta \Im{(\omega)} /\Im(\omega)_{\mathrm{A}.} \, _{\times 100 \%}$ \\    
        \hline  
        \texttt{mSI}$_\texttt{1024}$ & $0.419\,030$ & $0.418\,981\,(4)$ & $-0.011\,760\,(10)\%$\\   
        \texttt{PSI5}$_\texttt{1024}$ & $0.048\,589$ & $0.048\,727\,0\,(4)$ & $+0.284\,5\,(7)\%$\\
        \texttt{PSI10}$_\texttt{1024}$ & $0.048\,589$  & $0.048\,620\,4
        5\,(6)$ & $+0.065\,11\,(8)\%$\\
        \texttt{PSI20}$_\texttt{1024}$  & $0.048\,589$ & $0.048\,618\,46\,(23)$ & $+0.060\,99\,(3)\%$ \\
        \hline
        \texttt{PSI10}$_\texttt{512}$ & $0.048\,589$  & $0.048\,628\,4\,(6)$ & $+0.081\,35\,(9)\%$\\
        \hline
        \texttt{mSI} & $0.419\,030$ & $0.418\,966\,(4)$ & $-0.015\,150\,(13)\%$\\ 
        \texttt{PSI5} & $0.048\,589$  & $0.048\,745\,6\,(6)$ & $+0.322\,7\,(4)\%$\\
        \texttt{PSI10} & $0.048\,589$  & $0.048\,640\,2\,(8)$ & $+0.105\,69\,(18)\%$\\
        \texttt{PSI20} & $0.048\,589$  & $0.048\,638\,1\,(4)$ & $+0.101\,53\,(9)\%$\\
        \hline
        \texttt{PSI10}$_\texttt{disc.}$ & $0.048\,589$ & $0.043\,094\,(5)$ & $-11.308\,(12)\%$\\
        \texttt{PSI20}$_\texttt{disc.}$ & $0.048\,589$ & $0.047\,317\,6\,(11)$ & $-2.616\,(6)\%$\\
        \texttt{PSI40}$_\texttt{disc.}$ & $0.048\,589$ & $0.048\,395\,(7)$ & $-0.399\,(6)\%$\\
        \hline
    \end{tabular}
    \tablefoot{The error between the growth rate calculated analytically $\Im{(\omega)}_\mathrm{A.}$ and the growth rate from the numerical simulations $\Im{(\omega)}_\mathrm{N.}$ at $\mathbf{K} = \left(30,0,30\right)^T$.}
\end{table}

\begin{figure}
\centering
\includegraphics[width=\hsize]{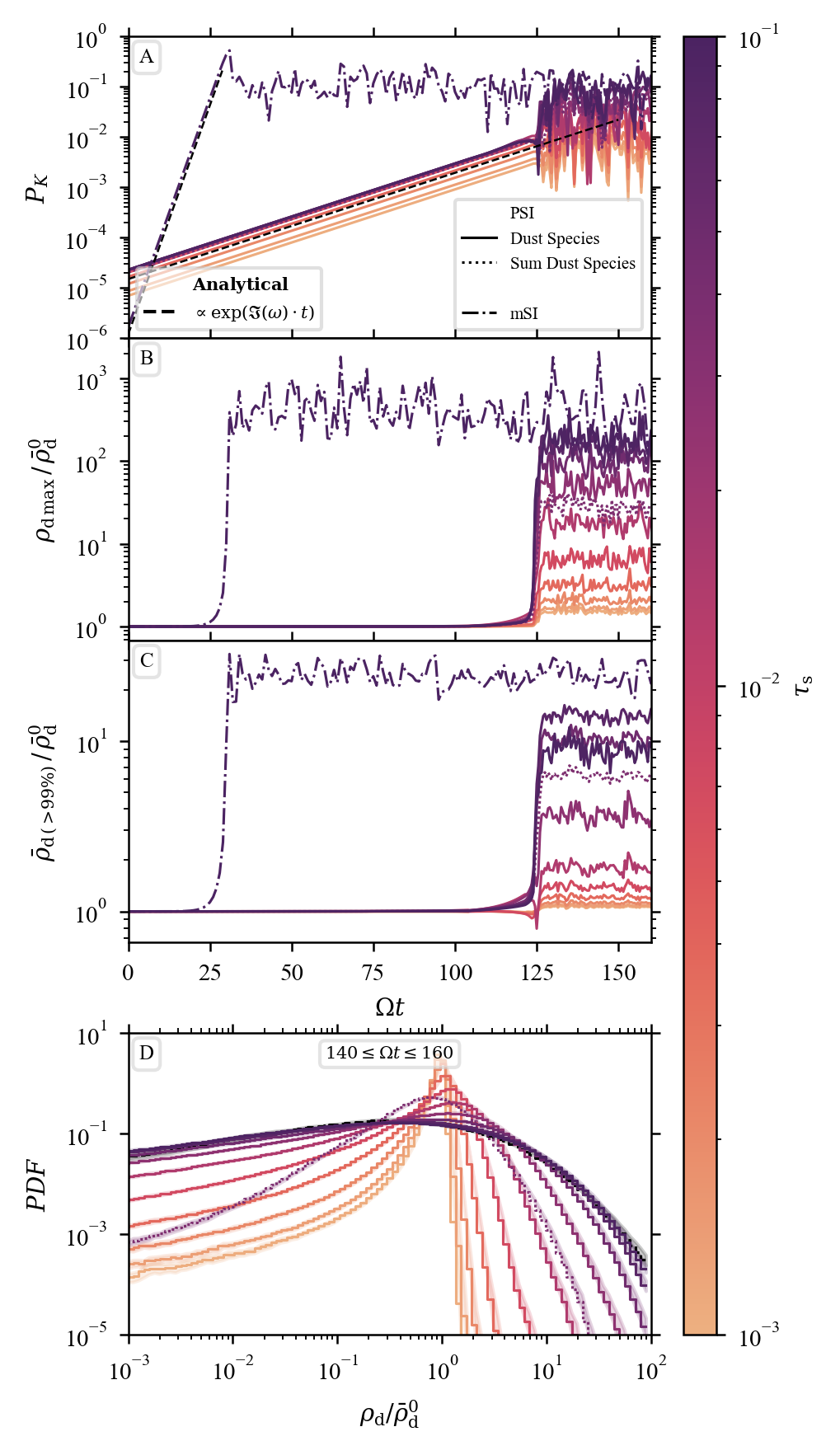}
  \caption{Time evolution of the mSI (run: \texttt{mSI}$_\texttt{1024}$) with dot-dashed line and PSI (run: \texttt{PSI10}$_\texttt{1024}$) with solid lines for the ten individual dust species and dotted line for the sum. Where plot A shows the amplitude of the largest mode in the shearing box (A sinusoid with wavenumber $\mathbf{K} = \left(30,0,30\right)^T$), plot B shows the maximum density in the shearing box. Plot C shows the mean density at every snapshot's $99^\mathrm{th}$ percentile. Plot D shows the average Probability Density Function (PDF) of the normalized density distribution in the nonlinear regime between ($140 \leq \Omega t \leq 160)$.}
     \label{fig:poly10}
\end{figure}

\par The perturbation will first grow exponentially \eqref{eq:pertubation}, where the growth rates of the SI are dependent on the wavenumber, diffusion and size distribution $\Im(\omega(\mathbf{K}, \alpha, \sigma))$. This growth has an analytical solution that can be calculated with software package \texttt{psitools} \citep{Paardekooper2020MNRAS.499.4223P, Paardekooper2021MNRAS.502.1579P, McNally2021MNRAS.502.1469M}. We compare the analytical growth rates with those from the numerical simulations. The analytical growth rate can be determined from the frequency $\omega$ \eqref{eq:pertubation}; for the polydisperse case $\omega_\texttt{PSI} = 0.42030787+ \mathrm{i} \ 0.04858883$ and for the monodisperse case $\omega_\texttt{mSI}= 0.34801869+ \mathrm{i} \ 0.4190302$. The time evolution of perturbations amplitude for a wave is shown in Figure \ref{fig:poly10}A. The amplitude is calculated with a Fast Fourier Transform \citep[with the software package \texttt{SciPy};][]{Virtanen2020NatMe..17..261V}. This amplitude can determine a growth rate corresponding to the numerical approach before it transitions into the nonlinear regime for our runs. This happens at $\Omega t \sim 125$ for PSI runs and $\Omega t \sim 30$ for the mSI runs. In Figure \ref{fig:poly10}, the coloured solid lines correspond to individual dust species at different Stokes numbers for the polydisperse case \texttt{PSI10}$_{1024}$, and the dotted line is the amplitude of the sum of the dust species. The monodisperse dust species at $\tau_\mathrm{s} = 0.1 \ \mathrm{\Omega^{-1}}$ is shown as a dot-dashed line. The growth rates and error between the analytical and numerical growth rates are given in Table \ref{tab:growthrate}. 

\begin{figure}
    \centering
    \includegraphics[width=\hsize]{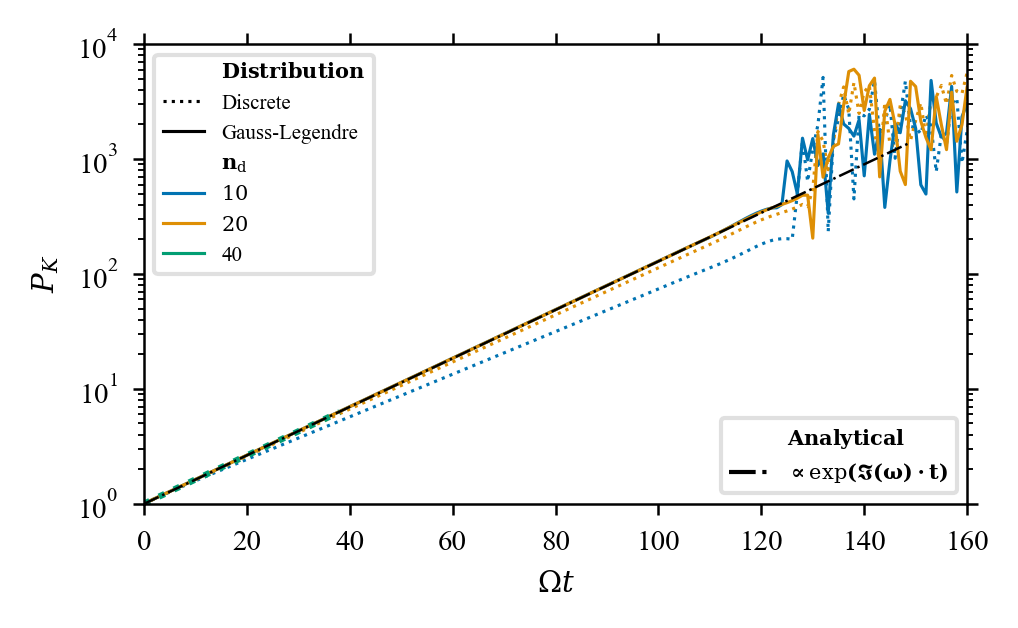}
    \caption{The normalized amplitude of the largest mode of the shearing box (a sinusoid with a wavenumber $\mathbf{K} = \left(30,0,30\right)^T$) for a different number of dust species (for blue $n_\mathrm{d}=10$, for orange $n_\mathrm{d}=20$ and for green $n_\mathrm{d}=40$) and different sampling method. The solid line indicates the GL method \eqref{eq:gausslegendre_tau}, and the dotted line indicates a log-linear sampling method (runs: \texttt{PSI20}, \texttt{PSI10}, \texttt{PSI40}$_\texttt{disc.}$, \texttt{PSI20}$_\texttt{disc.}$ and \texttt{PSI10}$_\texttt{disc.}$). The dot-dashed line is an analytical solution of the growth rate calculated with \texttt{psitools}.}
    \label{fig:sample_amp}
\end{figure}

\par In the linear regime, it is a lot easier to converge with spatial resolution than with the number of dust species \citep[see][]{Krapp2019ApJ...878L..30K, Zhu2021MNRAS.501..467Z}. The error in the growth rate is an indication of how good the sampling method approximates the continuum limit of the integrated momentum transfer between the gas and dust \eqref{eq:backreaction}. The error in the momentum transfer in the linear regime depends mainly on three parameters, the number of dust species used to sample the size distribution, the sampling method, and the spatial resolution. When we use the GL method, the error in the growth rate compared to the analytical value already converges between $n_\mathrm{d} = 5$ and $10$, see Table \ref{tab:growthrate}. This is also visible if we show the amplitude growth of the largest wavevector $\mathbf{K}=\left(30,0,30\right)$ between the GL method and the discrete method, see Figure \ref{fig:sample_amp}. In this Figure, the GL method (solid line) agrees with analytical results from \texttt{psitools} for both $n_\mathrm{d}=10$ and $20$. While the amplitude of the discrete method (dotted line) also converges with the analytical results, they are significantly less accurate than the GL method at the same number of dust species. Using the discrete sampling method, the error in the growth rate at $n_\mathrm{d}=40$ is similar to the error using the GL method when using only $n_\mathrm{d}=5$ (Table \ref{tab:growthrate}), which is a lot less computationally expensive to run.

\begin{figure}
\centering
\includegraphics[width=\hsize]{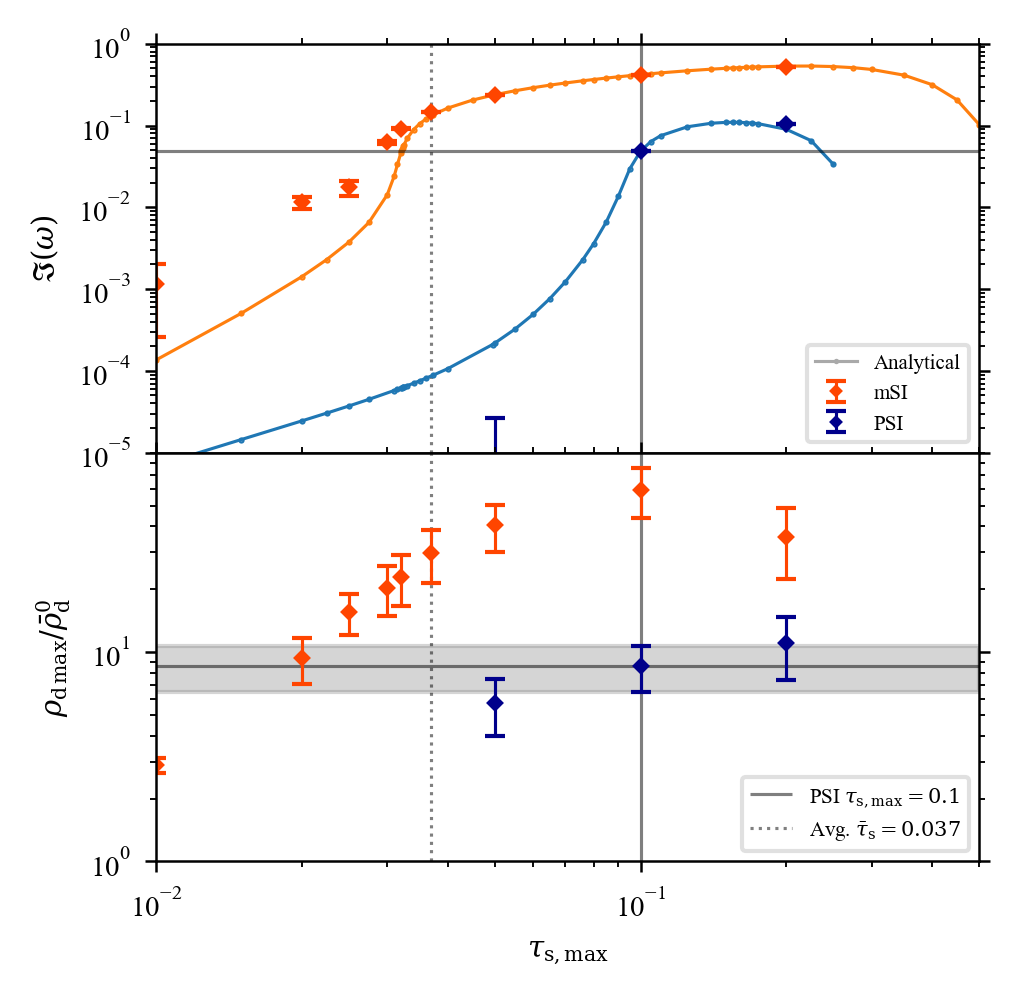}
  \caption{The top plot shows the growth rate in the linear regime at wavevector $\mathbf{K}=\left(30,0,30\right)^T$ for setups with different maximum Stokes numbers $\tau_\mathrm{s, max}$, the solid orange and blue line are the analytical growth rates (for mSI and PSI, respectfully) calculated with \texttt{psitools} \citep{McNally2021MNRAS.502.1469M}. The diamonds indicate the fitted values of the growth rate in the linear regime of the numerical simulations. In the bottom plot, the diamonds correspond to the fitted values of the saturated amplification factor of the maximum density in the non-linear regime. The solid grey lines indicate the values for run \texttt{PSI10}, and the dotted grey line indicates the average Stokes number of the size distribution of run \texttt{PSI10}.}
     \label{fig:max_stokes_number}
\end{figure}

The growth rate is also dependent on the maximum Stokes number, this is visualized in the top plot in Figure \ref{fig:max_stokes_number}. The analytical growth rates calculated with \texttt{psitools} are visualized with the solid lines. We see that the growth rates decrease for lower Stokes numbers and are always lower for PSI than mSI. Even if we compare the growth rate of the mSI at the average Stokes number of a PSI size distribution, the growth rate of PSI is still smaller. For the specific example of the PSI with $\tau_\mathrm{s, \, max}=0.1 \Omega^{-1}$, which has an average Stokes number $\bar{\tau}_\mathrm{s} = 0.037 \, \Omega^{-1}$, the mSI growth rate is still $2.803$ times higher than the growth rate of the PSI. The Stokes number of the mSI would have to be $\tau_\mathrm{s} = 0.0321 \, \Omega^{-1}$ to have a similar growth rate as the PSI.

\par The fitted values of the growth rates of the numerical runs agree well with the analytical growth rates in the top plot of Figure \ref{fig:max_stokes_number} for $\tau_\mathrm{s, \ max} \gtrsim 0.035  \, \Omega^{-1}$. For lower $\tau_\mathrm{s}$, however, we see the fitted mSI growth rates (red diamonds), start to deviate from the analytical growth rates (orange line). This can be explained by faster growing high-$K$ modes; these get excited by small numerical errors and can boost the power $P(K)$ at wavenumber $\mathbf{K} = \left(30,0,30\right)^T$ and can even become the largest perturbed mode before the instability reaches the nonlinear regime. This happens for runs \texttt{mSI}$_{\tau_\mathrm{s} \leq 0.035}$, run 
\texttt{PSI}$_{\tau_\mathrm{s}, \, 5\mathrm{e}-2}$, and run \texttt{PSI}$_{\beta, -3.8}$. This is also visible in the time evolution of the amplitude, where the amplitude suddenly increases a lot faster, like the blue line in the top plot of Figure \ref{fig:ts_peak_dens}.

\section{Nonlinear regime}\label{sec:non_linear}

When the amplitude of the perturbation gets high enough, the wave will break and transition to a turbulent nonlinear state, showing vortices-like motion in the x,z-plane. Even though the non-linear regime is not analytically tractable, the fact that it saturates allows us to quantify it statistically; therefore, we can extract trends from this regime. The important feature is that the non-linear regime will produce high-density structures. This is also important for planet formation because the high-density structure can lead to clumping when we include self-gravity, and then the dust clumps can collapse into planetesimals when the Roche density is exceeded \citep[see, e.g. the strong clumping criteria of][]{Li2021ApJ...919..107L}. The extent of the trapping in the monodisperse case depends on the spatial resolution and Stokes number because we consider the dust to be a pressureless fluid with no diffusion. The spatial resolution in a grid-based hydrodynamical code determines the minimum size of structures that can be resolved. This also puts a constraint on the size of the clump, and larger clumps will have smaller maximum densities; the convergence of spatial resolution is further discussed in Section \ref{sec:resolution}. 

\begin{figure*}
    \includegraphics[width=\hsize]{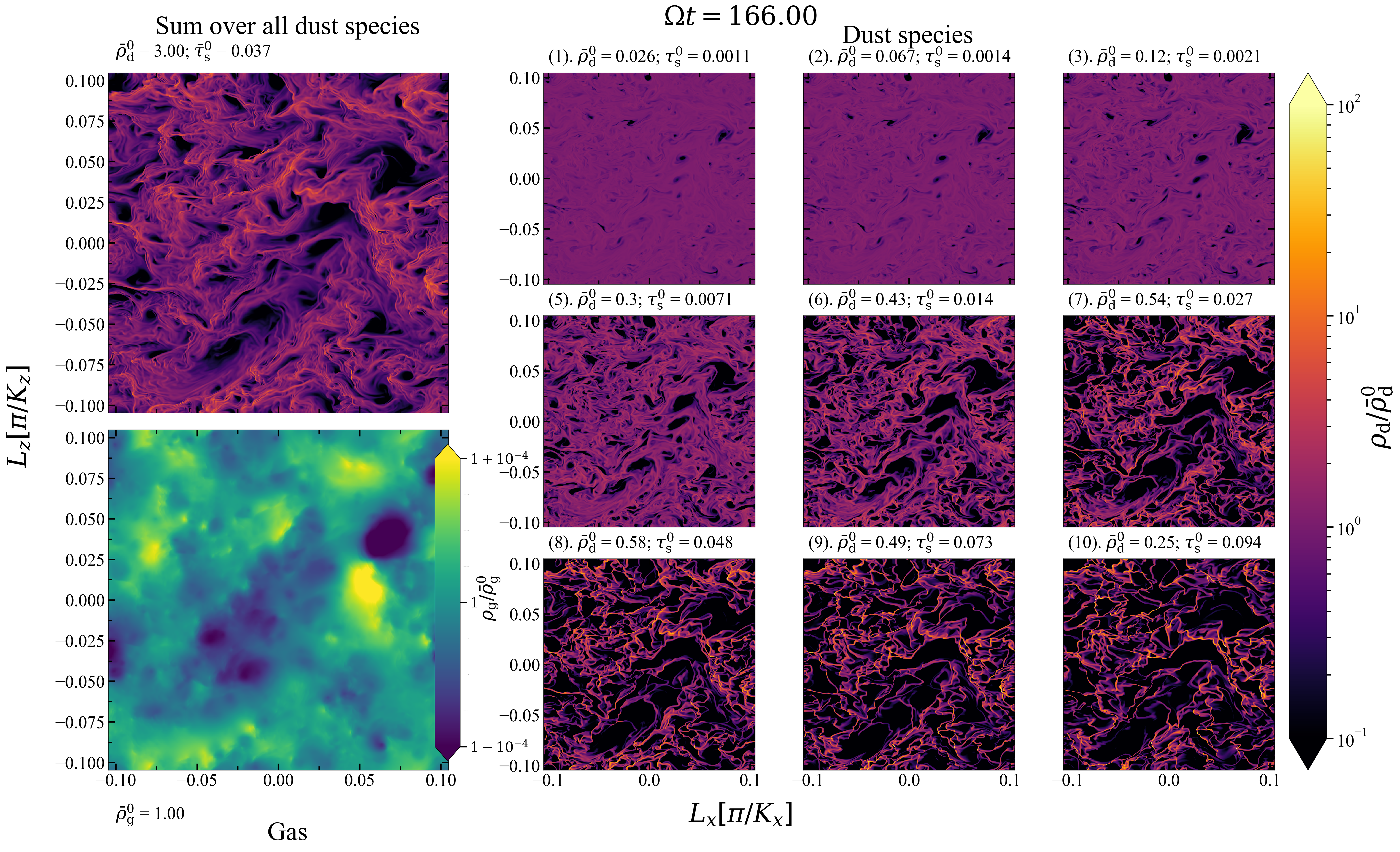}
    \caption{A snapshot of the normalized density in the nonlinear regime ($\Omega t=166$), for the PSI run \texttt{PSI10}$_\texttt{1024}$, showing the density of the sum of the dust species (upper left), gas (lower left) and nine of the ten individual dust species with increasing Stokes number.)}
    \label{fig:normalized_density}
\end{figure*}
   
\par We expect higher dust densities for species with higher Stokes numbers since smaller dust is strongly coupled to the gas, which in turn is (nearly) incompressible. This means that the density distribution for smaller dust sizes will be more homogeneous, affecting the amplification factor for monodisperse distributions, see the bottom plot in Figure \ref{fig:max_stokes_number}. In this plot, we can see that the amplification factor of the maximum density becomes lower for lower Stokes numbers. This is also the case for the individual dust species in PSI runs, where the smaller dust sizes are distributed more homogeneously. In figure \ref{fig:normalized_density}, we look at a snapshot of the normalized density for all the combined dust and the different dust sizes of run \texttt{PSI10}$_{1024}$, where we can see that the dust sizes are correlated; low-density and high-density regions occur in the same place for different dust species. The figure also shows that the amplification factor is lower for the smaller dust species. The normalized density for the largest dust bins shows more structure than the smallest species and is also visible if we plot the time evolution of the maximum density and the time-averaged density distributions, see Figure \ref{fig:poly10} B and D, respectively. In  Figure \ref{fig:poly10}B, the dot-dashed line shows the monodisperse case from run \texttt{mSI}$_\texttt{1024}$ and the solid line the individual dust species, where the colour indicates the Stokes number and the average Stokes number of the sum of the size distribution. In this figure, we can see that the amplification factor for mSI is largest at $484 \pm 324$. The amplification factor of the PSI is significantly lower at $27.3\pm4.6$.

\par Considering the strong $\tau_s$ dependence of the SI, comparing monodisperse and polydisperse simulations with a monodisperse $\tau_s$ equal to the maximum of the dust distribution of the corresponding polydisperse distribution is expected to yield higher dust densities in the monodisperse case, see bottom panel of Figure \ref{fig:max_stokes_number}. Correcting for this bias, comparing polydisperse simulations to monodisperse ones with a $\tau_s$ equal to that of the average of the size distribution still shows higher dust densities in the monodisperse case. For the specific example with run \texttt{PSI10}, the average Stokes number is $\bar{\tau}_\mathrm{s} = 0.037 \, \Omega^{-1}$. At this Stokes number, the mSI amplification factor of the maximum density is still $3.5$ times higher than the amplification factor of run \texttt{PSI10}. The Stokes number of the mSI would have to be $\tau_\mathrm{s} = 0.02 \, \Omega^{-1}$ to have a similar amplification factor. This means that a PSI can \emph{not} be represented by an mSI run with properties that are derived from the averaged properties of the PSI setup, like average Stokes number and total dust density of the size distribution.

\par The PSI is not a simple summation of individual discrete mSI cases at different Stokes numbers. One way the PSI differs from mSI is in terms of growth rates. The growth rates of all the individual dust species are identical for the PSI case, but this depends on the Stokes numbers in the mSI case, see Section \ref{sec:stokesrange}. This also translates into the transition time of the individual dust species of the PSI. They all transition simultaneously (see Figure \ref{fig:poly10} A-C), whilst the transition time mSI changes with Stokes numbers (however, the shared point of transition for PSI is also different for different Stokes ranges, see Section \ref{sec:stokesrange}). There are also differences between discrete mSI cases and individual dust species of the PSI. The amplification factor is lower than the mSI case, and if (see Figure \ref{fig:normalized_density}). The dust sizes are correlated; low-density and high-density regions occur in the same place for different dust species.

\subsection{The size distribution in the clumps} \label{sec:evolution_sizedist}

\par In Figure \ref{fig:poly10}B, we can see that when considering the maximum dust density, the amplification factor increases monotonically with $\tau_s$, as expected from linear theory and non-linear monodisperse simulations. However, in Figure \ref{fig:poly10}C, we plot the mean of the top $1\%$ of the dust density. We see that the trend is subtly different: the dust density increases with $\tau_s$ for most of the Stokes number range, but the maximum density occurs at $\tau_s$, slightly smaller than the maximum. Since the top $1\%$ of the dust density probes more extended structures than the absolute maximum dust density, this indicates that there is a peak in the size distribution in the clumps.

\begin{figure}
    \centering
    \includegraphics[width=\hsize]{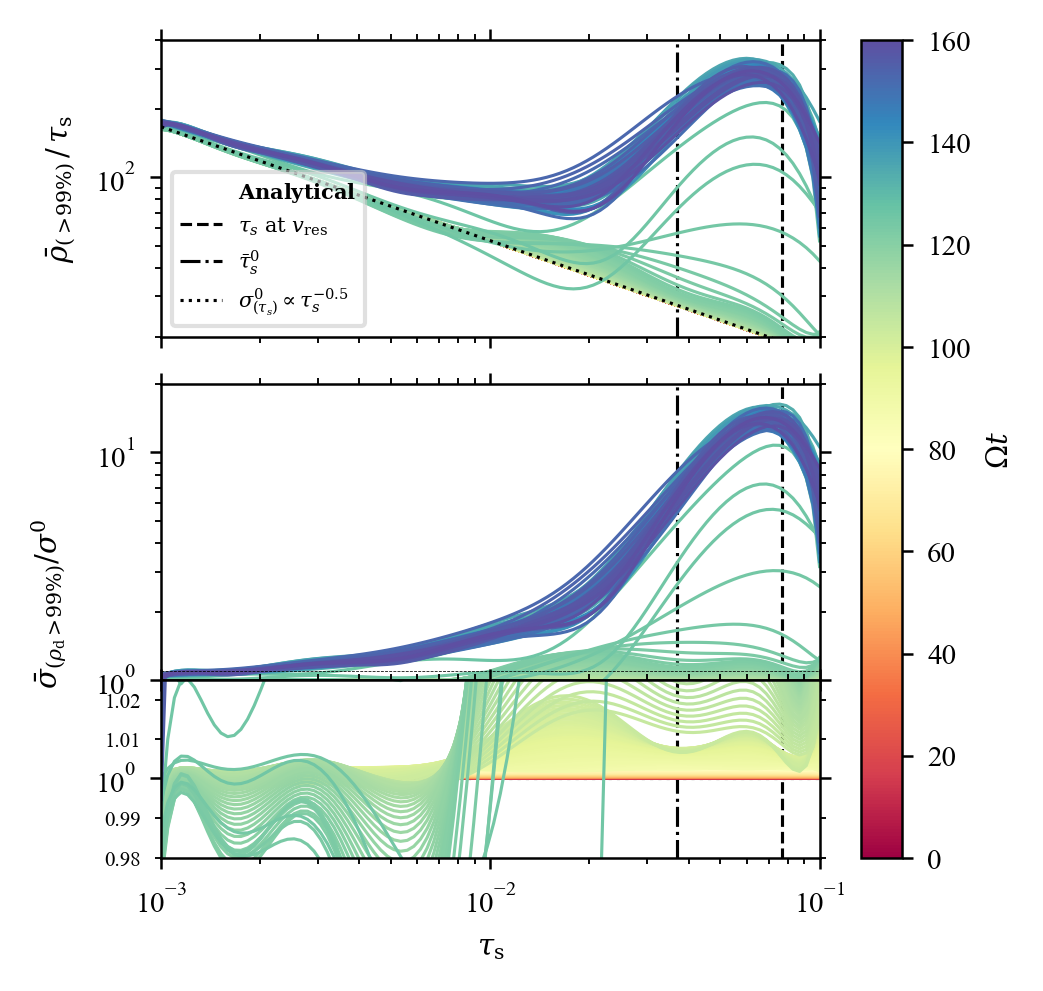}
    \caption{Top: The mean size distribution in the $99^\mathrm{th}$ percentile at different times, with the dotted line indicating the size distribution at $\Omega t =0$. The bottom shows the same size distribution normalized by the background-size distribution $\sigma^0(\tau_\mathrm{s})$. The dashed line indicates the Stokes number corresponding to the resonant velocity, and the dot-dashed line is the average Stokes number of the size distribution at $\Omega t=0$.}
    \label{fig:sizedistribution}
\end{figure}

\par We can better visualize this peak in the size distribution by plotting the size distribution $\sigma(\tau_\mathrm{s})$. In Figure \ref{fig:sizedistribution}, we show the time evolution of a polynomial that passes through the GL points of PSI simulation (from run \texttt{PSI10}$_\texttt{1024}$) throughout the simulation.  The top panel of Figure \ref{fig:sizedistribution} shows size distribution at the $99^\mathrm{th}$ percentile, with the dotted line indicating the MRN size distribution at the start of the simulation. The bottom panel shows the same size distribution normalized by the background distribution and has a subplot zoomed at smaller densities. 

\par The size distribution in the nonlinear regime ($\Omega t > 125$) settles into a trend with increasing densities at higher $\tau_\mathrm{s}$, but shows a clear peak. The peak in the size distribution could indicate either a resonance corresponding to the drift velocity or an inherent dependence on the Stokes range. From the mSI theory of the RDI, we can expect a strong response at size densities where the radial phase velocity of the perturbation matches the background radial drift velocity at a certain Stokes number. 
\begin{align}
    \frac{\Re(\omega)}{k_x} = w^0_{\mathrm{s,}\, x}
    \label{eq:resonance}
\end{align}
The PSI does not have this simple response with size distribution because of the distribution of dust sizes. There is a range of radial drift speeds $u_x$ (see Eq. \ref{eq:dustx_equilibrium_solution}) dependent on the dust size, whilst the resonance can only be satisfied by a single velocity. However, we can use equation \eqref{eq:dust_momentum} and \eqref{eq:dustx_equilibrium_solution} to find the mean radial drift speed $\bar{u}^0_x$ of the size distribution. The mean drift velocity can be compared to the mSI drift speed ${v_\mathrm{d}}^0_x(\tau_\mathrm{s})$ where we define a resonance size $\tau_\mathrm{s}$ where ${v_\mathrm{d}}^0_x(\tau_\mathrm{s}) = \bar{u}^0_x(\sigma)$ holds \citep[from][]{Paardekooper2021MNRAS.502.1579P}. In Figure \ref{fig:sizedistribution}, the Stokes number corresponding to this response is indicated with a vertical dashed line, and the average Stokes number $\bar{\tau}^0_s$ is indicated with the dot-dashed line. The peak of the size distribution visually overlaps with the resonance size but does not convincingly correspond to it. On the other hand, the peak in the size distribution can be an inherent feature of the Stokes number range and is affected by the maximum Stokes number $\tau_\mathrm{s, \, max}$.

\subsection{Convergence} \label{sec:convergence}

\par It is also possible that the peak in the size distribution corresponds to a numerical error. If this is the case, the peak should depend on the spatial resolution or the number of dust species used to sample the continuum size distribution. Another possibility is that the polynomial interpolation through the density of the individual dust species at the Stokes numbers that are sampled through the GL points will appear to indicate a peak, whilst the densities of the discrete dust species do not indicate this. We can validate this by using a larger number of dust species as well as comparing the GL method to the discrete method.

\subsubsection{Spatial resolution} \label{sec:resolution}

\begin{figure}
    \centering
    \includegraphics[width=\hsize]{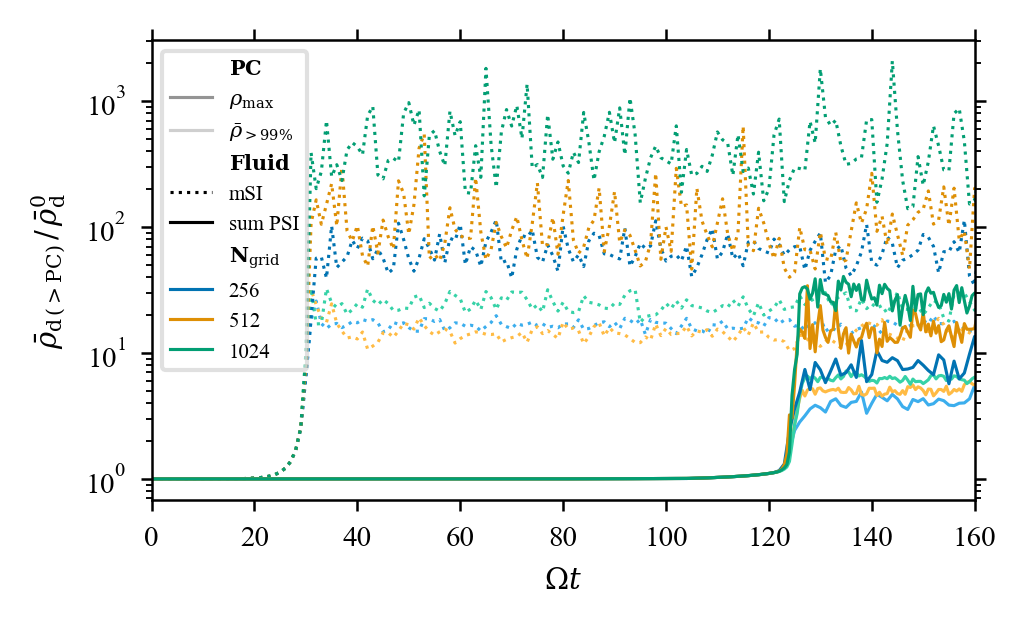}
    \caption{The normalized (mean) density at densest pixel and at $99^\mathrm{th}$ percentile (indicated by darker and lighter colours; respectively.) for different spatial resolutions, the mSI runs at different resolutions are indicated with the dotted line for run \texttt{mSI}$_\texttt{1024}$, \texttt{mSI}$_\texttt{512}$ and  \texttt{mSI} using green, orange and blue respectively and the PSI runs \texttt{PSI10}$_\texttt{1024}$, \texttt{PSI10}$_\texttt{512}$ and  \texttt{PSI10} with the same colour coding as mSI.}
    \label{fig:resolution_density}
\end{figure}

\par The spatial resolution directly limits the smallest structures that can be resolved within the simulation and the maximum density that is possible to create through clumping. Together with the fact that the dust fluid is fully compressible (unlike the gas), the simulations will never converge with spatial resolution unless dust diffusion is implemented, see Section \ref{sec:diffusion}. The dust without diffusion can be compressed into arbitrarily small volumes, and the maximum density in the nonlinear regime will increase with increasing spatial resolution. This increase in density is not as strong if we look at the mean of larger percentiles. This is visible in Figure \ref{fig:resolution_density}, showing the amplification factor for the mSI and PSI simulations at a spatial resolution of $256$, $512$ and $1024$. For the PSI runs, the saturated mean density at the $99^\mathrm{th}$ percentile and maximum density increases with resolution, although the difference between resolutions is less at the $99^\mathrm{th}$ because we are analyzing larger structures. The linear regime of the PSI and mSI is not affected by the resolution, and the transition from the linear regime to the nonlinear regime happens at the same time for different spatial resolutions, indicating that the linear regime already fully converged before a spatial resolution of $256\times256$ (Table \ref{tab:growthrate}). The amplification factor of the $99^\text{th}$ percentile in the saturated regime for a resolution of \texttt{mSI}$_{512}$ is lower than at a resolution of \texttt{mSI}$_{256}$, this is counter-intuitive but not significant. The error is smaller than the $1-\sigma$ variation of density, and the maximum density of \texttt{mSI}$_{512}$ is higher than \texttt{mSI} at a $N_\mathrm{grid} = 256$\footnote{In the mSI runs, we observe a numerical artefact in the gas density on the scale of the grid cells, the PSI is unaffected. A similar error in the mSI is also mentioned in Section 3.5.5 of \citet{Benitez-Llambay2019ApJS..241...25B}.}.

\begin{figure}
    \centering
    \includegraphics[width=\hsize]{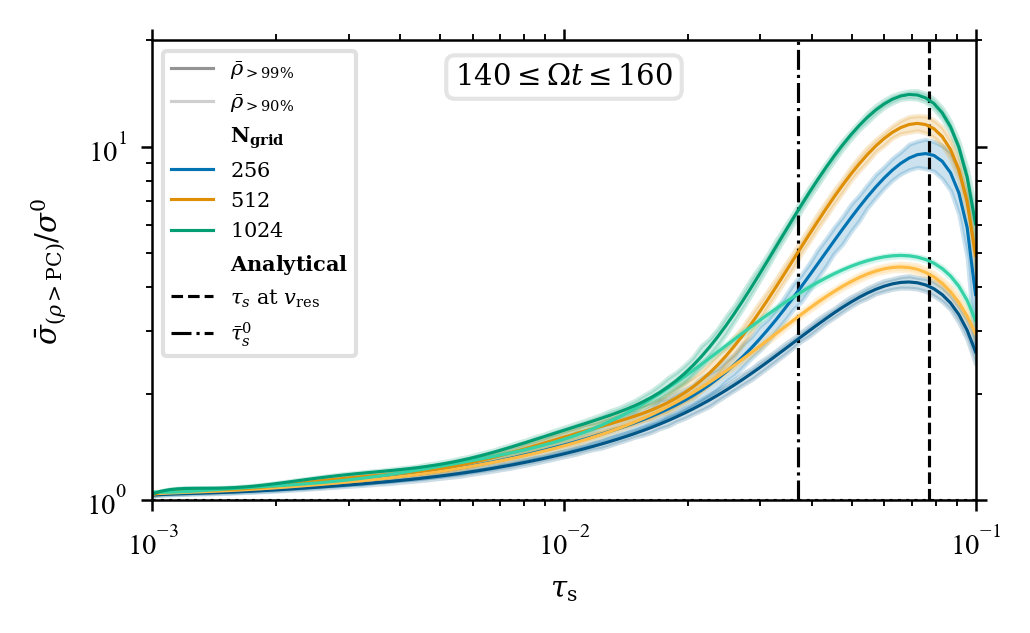}
    \caption{The normalized mean size distribution at the $99^\mathrm{th}$ and $90^\mathrm{th}$ percentile and between $140 \! \leq \! \Omega t \! \leq \! 160$ using the same PSI runs and colour scheme as Figure \ref{fig:resolution_density}.}
    \label{fig:resolution_sizedist}
\end{figure}

\par The overall shape of the peak in the size distribution in the highest density regions is consistent between resolutions, shown in Figure \ref{fig:resolution_sizedist}. The height of the size distribution increases slightly at higher Stokes numbers. At higher resolutions, the maximum amplification factor increases; therefore, $99^\mathrm{th}$ percentile covers less area, decreasing the relative presence of smaller Stokes numbers that are more homogeneously distributed around the shearing box. This is also why the size distribution is flatter when we average over the  $90^\mathrm{th}$ percentile (also shown in Figure \ref{fig:resolution_sizedist}). We average over a larger area, where the smaller, more homogeneous dust species are relatively more abundant compared to the higher Stokes numbers that are more clumped. The simulations follow the relation between the overabundance of larger Stokes numbers in the highest-density regions and spatial resolution, but the peak (underrepresentation of the largest Stokes numbers) also stays consistent with different resolutions. There is a weak trend in the position of the peak to drift to smaller stroke numbers with higher resolution, although the change in Stokes number is a lot smaller than the difference between different-sized dust species.

\subsubsection{Number of dust species}\label{sec:ndustspecies}

\begin{figure}
    \centering
    \includegraphics[width=\hsize]{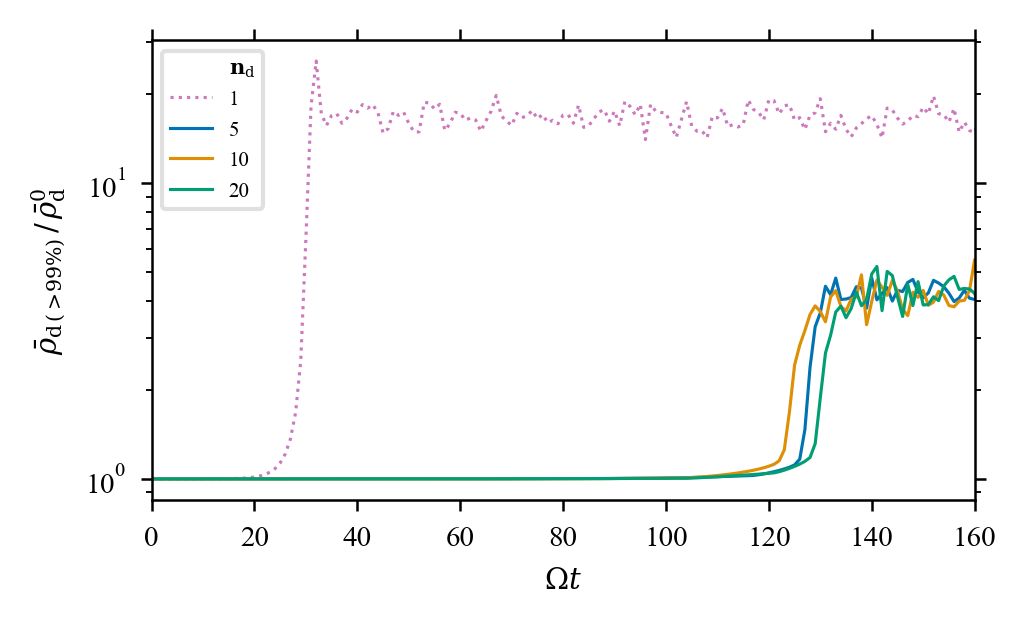}
    \caption{The normalized mean density at $99^\mathrm{th}$ percentile for different numbers of dust species from run  \texttt{mSI}$_\texttt{1024}$ indicated with a dotted pink line, run \texttt{PSI5}$_\texttt{1024}$ in blue, run \texttt{PSI10}$_\texttt{1024}$ in orange and run \texttt{PSI20}$_\texttt{1024}$ in green.}
    \label{fig:ndust_density}
\end{figure}

\par There are no big differences between simulations at different numbers of dust species in the linear regime (Section \ref{sec:linear}), but Figure \ref{fig:ndust_density} does show that the transition point from the linear regime to the nonlinear regime shifts slightly with $n_\mathrm{d} =10$ being slightly earlier and $n_\mathrm{d}=20$ slightly later. The fact that the smallest sample rate is in the middle does not indicate any clear trend, but this can also be that $n_\mathrm{d}=5$ is still far from converging. Because the saturation level between $n_\mathrm{d} =10$ and $n_\mathrm{d} =20$ is similar, whilst the saturated amplification factor of $n_\mathrm{d}=5$ is significantly higher. 

\begin{figure}
    \centering
    \includegraphics[width=\hsize]{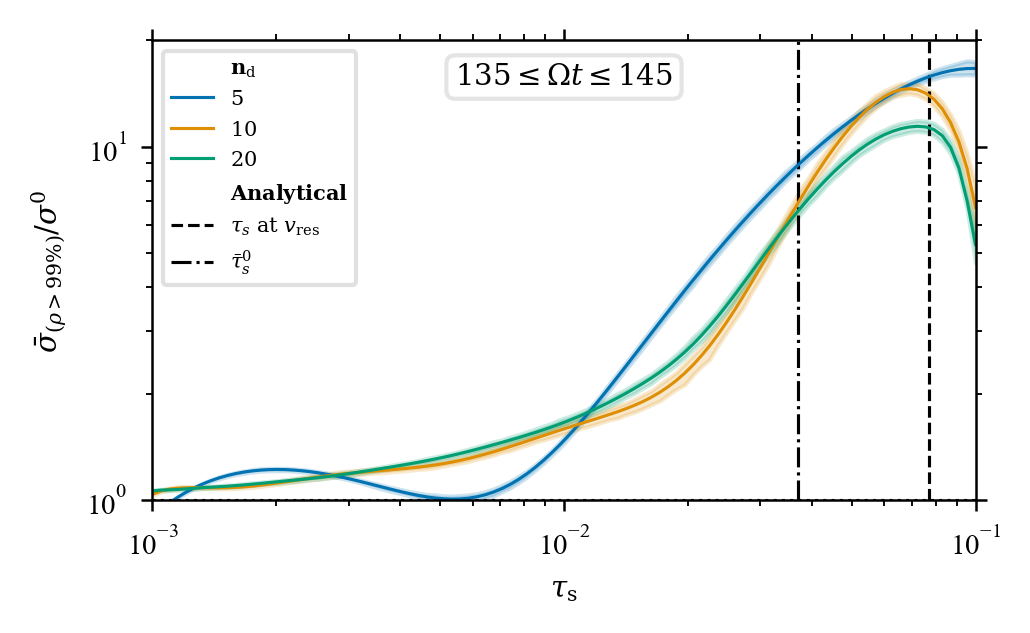}
    \caption{The normalized mean size distribution at the $99^\mathrm{th}$ percentile and between $135 \! \leq \! \Omega t \! \leq \! 145$ using the same PSI runs and colour scheme as Figure \ref{fig:ndust_density}. The dashed line indicates the Stokes number at resonance velocity  ${v_\mathrm{d}}^0_x(\tau_\mathrm{s}) = \bar{u}^0_x(\sigma)$ and dot-dashed line the average Stokes number at $\Omega t=0$.}
    \label{fig:ndust_sizedist}
\end{figure}

\par Figure \ref{fig:ndust_sizedist} shows that mean size distribution in the $99^\mathrm{th}$ percentile does not show any clear correlation between dust sizes for $n_\mathrm{d}=5$ except an increased abundance of larger Stokes numbers. There is a clear peak in the size distribution at $n_\mathrm{d}=10$ and $n_\mathrm{d}=20$, where the peaks appear to be more skewed at a higher sampling rate of $n_\mathrm{d}=20$ compared to $n_\mathrm{d}=10$. Although this sharper drop at the largest Stokes numbers does not appear for $n_\mathrm{d}=20$ at a lower spatial resolution of $256\times256$, see Figure \ref{fig:sample_sizedistribution}. The size distribution remains smooth in size space for an increasing number of dust species, indicating that the GL method is also a good approximation for the continuum in the non-linear regime.

\subsubsection{Sampling method}\label{sec:sample_method}

\par As discussed in Section \ref{sec:gaussianlegendrequadrature}, the GL approximates the integral for the impulse transfer between the dust species and the gas \eqref{eq:backreaction} is different from the uniform sampling method in logspace, that we defined as the discrete method. The error between the continuous limit and discrete integral decreases faster with the number of dust species using the GL method than using the discrete method, see Section \ref{sec:linear}. 

\begin{figure}
    \centering
    \includegraphics[width=\hsize]{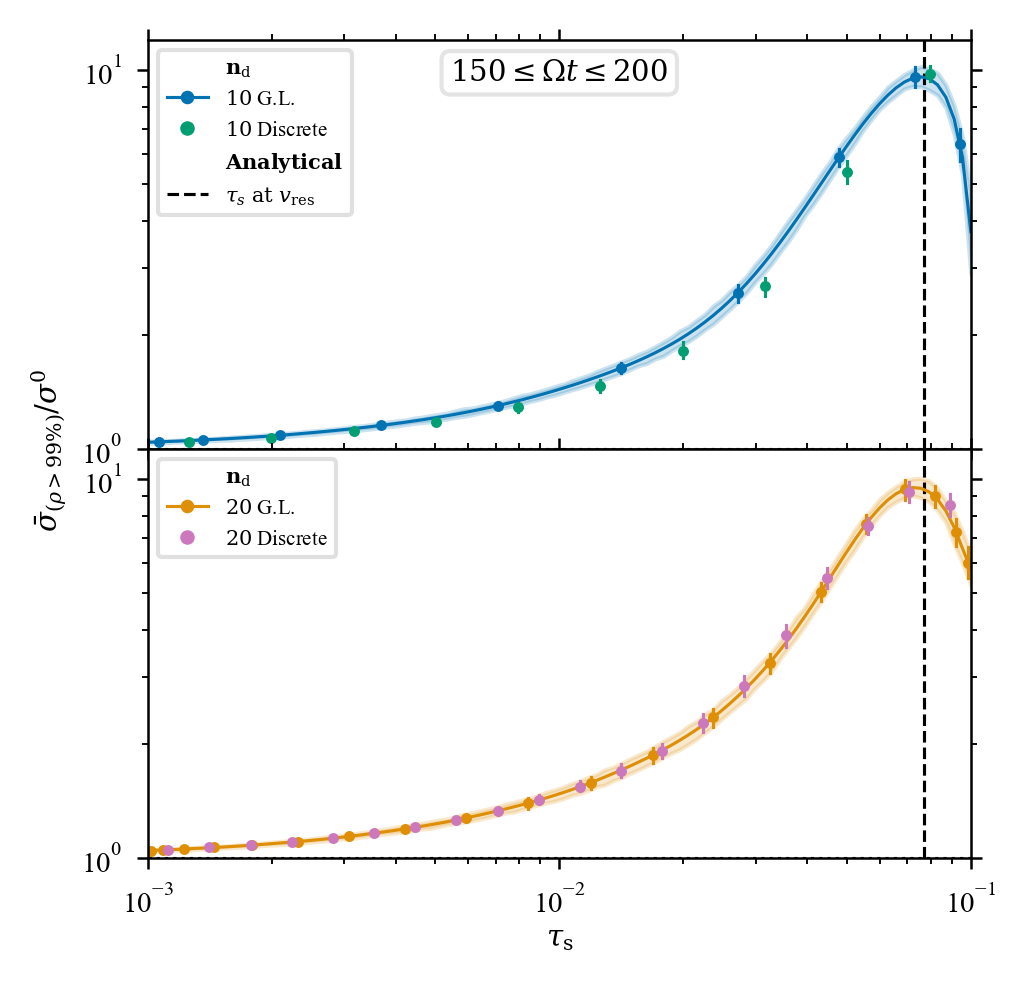}
    \caption{The normalized mean size distribution at the $99^\mathrm{th}$ percentile and between $150 \! \leq \! \Omega t \! \leq \! 200$. Where the top plot shows the PSI run with $10$ dust species sampled from the GL in blue (run  \texttt{PSI10}) and the PSI run with $10$ dust species uniformly sampled from a logarithmic scale in green (run  \texttt{PSI10}$_\texttt{disc.}$) and the bottom plot shows the PSI run using 20 dust species sampled from the GL in orange (run  \texttt{PSI20}) and the PSI run with 20 dust species uniformly sampled from a logarithmic scale in pink (run \texttt{PSI20}$_\texttt{disc.}$).}
    \label{fig:sample_sizedistribution}
\end{figure}

The effect of the GL sampling method in the nonlinear regime is less straightforward because the local size distribution can change significantly. It is possible that the size distribution can become more monodisperse locally and that the back reaction on the gas will be dominated by only part of the size distribution. This could affect the integration error. It is also possible that the peak in the size distribution in the densest regions could be an artefact of the weights or location of the roots of the polynomial or the projection of roots  $x_i$ between $[-1;1]$ to the Stokes numbers $\tau_{s, i}$ within the Stokes range $[\tau_{s,\mathrm{min}};\tau_\mathrm{s, \, max}]$, given by \eqref{eq:gausslegendre_tau} and \eqref{eq:gausslegendre_rho}. Therefore, we reconstructed the size distribution of the simulation from the discrete method (run \texttt{PSI10}$_\texttt{disc.}$ and \texttt{PSI20}$_\texttt{disc.}$) and compare it to the GL method, shown in Figure \ref{fig:sample_sizedistribution}. 

\par In the case of $10$ dust species (Figure \ref{fig:sample_sizedistribution} top panel), the slight offset between the values of Stokes numbers of the two sampling methods means that the discrete sampling (shown in green) of the individual dust bins does not sample the size distribution at Stokes numbers larger than the peak. Therefore, we can not conclude if there is a peak in the size distribution by looking at $n_\mathrm{d} =10$, but the discrete method does follow the same general trend of the GL sampling method and are not in disagreement. 

\par At a higher dust sampling rate of $n_\mathrm{d} = 20$ (Figure \ref{fig:sample_sizedistribution} bottom panel), where the discrete distribution does sample Stokes numbers larger than the peak, we see at least a reduced presence of the largest Stokes number that is consistent with the peak in the size distribution seen in all the simulation runs using the GL sampling method. This shows that the peak in the size distribution in the densest regions is physical rather than a numerical artefact.

\section{Parameter study}\label{sec:parameterstudy}

To get a better indication of where the peak of the size distribution comes from and what can influence its location, we did a parameter study where we covered different values for the dust diffusion coefficient $\alpha$, maximum Stokes number $\tau_\mathrm{s, \, max}$, the slope of the size distribution $\beta$ and the dust-to-gas ratio $\mu$, see Table \ref{tab:simulations}. This parameter study is done at a fixed spatial resolution of $\text{N}_\text{grid}=256$ and using $10$ dust species for the polydisperse simulations. 

\subsection{Diffusion}\label{sec:diffusion}

\begin{figure}
    \centering
    \includegraphics[width=1\linewidth]{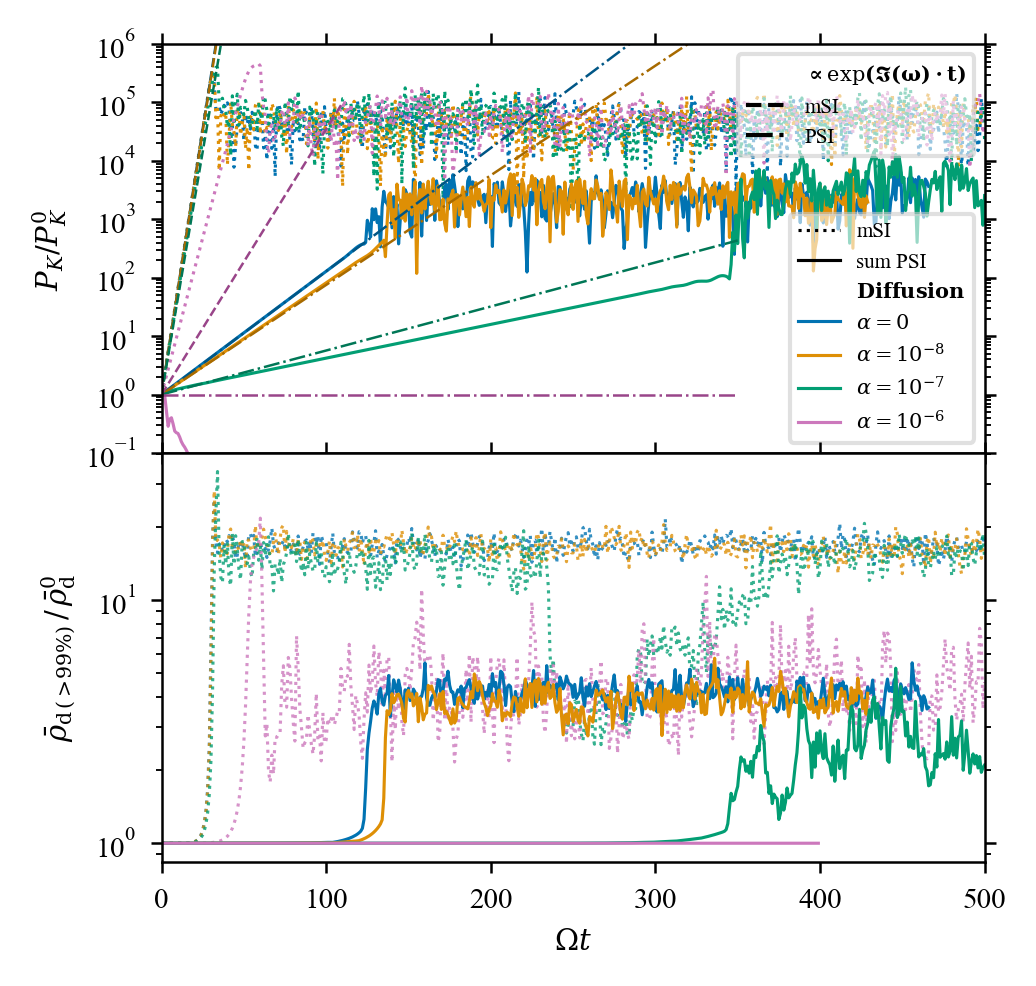}
    \caption{Time evolution of the mSI (runs: $\left[\texttt{mSI}^*, \texttt{mSI}_{\alpha,1\mathrm{e}-8}, \texttt{mSI}_{\alpha,1\mathrm{e}-7}, \texttt{mSI}_{\alpha,1\mathrm{e}-6}\right]$) with dotted line and PSI (runs: $\left[\texttt{PSI}^*, \texttt{PSI}_{\alpha,1\mathrm{e}-8}, \texttt{PSI}_{\alpha,1\mathrm{e}-7}, \texttt{PSI}_{\alpha,1\mathrm{e}-6}\right]$ ) with solid lines for different dust diffusion coefficients $\alpha$. Where the top plot shows the amplitude of the largest mode in the shearing box (A sinusoid with wavenumber $K_{x,z}$ = 30), and the dashed lines and dash-dotted lines showing the analytical results calculated with \texttt{psitools} for the corresponding $\alpha$. The bottom plot shows the normalized mean density at the $99^\text{th}$ percentile for every snapshot.}
    \label{fig:diffusion_time}
\end{figure}

The diffusion coefficient $\alpha$ is a dimensionless parameter that quantifies the strength of the turbulence in protoplanetary discs and the viscosity $\nu = \alpha c_\mathrm{s} H.$ \citep{Shakura1973A&A....24..337S}. We used the $\alpha$ coefficient for the gas and dust diffusion in \texttt{FARGO3D}. The dust diffusion is modelled with a continuity equation for (pressureless) dust fluids, spreading mass depending on the gradient of the concentration (see appendix of \citet{Weber2019ApJ...884..178W} for the implementation in \texttt{FARGO3D}).

\par Turbulence in a protoplanetary disc stirs the dust and can work to concentrate or disperse it, making it possible to form local regions of higher dust densities but also to destroy the formed substructure and clumps \citep{Johansen2007Natur.448.1022J, Yang2018ApJ...868...27Y, Schafer2020A&A...635A.190S, Lim_a2024ApJ...969..130L}. The diffusion coefficient is important for addressing the finite resolution of hydrodynamical grid codes, where implementing a dust diffusion model can be a solution to the infinite compressibility of dust in the simulations and the unresolved turbulence. An important caveat is that the alpha model is limited on the small scales and can \emph{not} capture the turbulence concentrating effect. 

\par In the linear regime, the growth rates get smaller when the diffusion coefficient increases for the mSI  \citep[see analytical work of][]{Youdin2005ApJ...620..459Y, Umurhan2020ApJ...895....4U, Chen2020ApJ...891..132C} and for PSI \citep[see][]{McNally2021MNRAS.502.1469M}. The growth rates are shown in the top plot of Figure \ref{fig:diffusion_time}, for runs $\left[\texttt{mSI}_{\alpha,1\mathrm{e}-8}, \texttt{mSI}_{\alpha,1\mathrm{e}-7}, \texttt{mSI}_{\alpha,1\mathrm{e}-6}, \texttt{PSI}_{\alpha,1\mathrm{e}-8}, \texttt{PSI}_{\alpha,1\mathrm{e}-7}, \texttt{PSI}_{\alpha,1\mathrm{e}-6}\right]$ where for the last polydisperse run with $\alpha = 10^{-6}$ there is no exponential growth and PSI does not develop. The lower plot shows that diffusion also affects the amplification factor of the SI, lowering the amplification factor for higher values of diffusion coefficients. The densities corresponding to the $99^\text{th}$ percentile are also more variable in time at higher diffusion coefficients.

\begin{figure}
    \centering
    \includegraphics[width=1\linewidth]{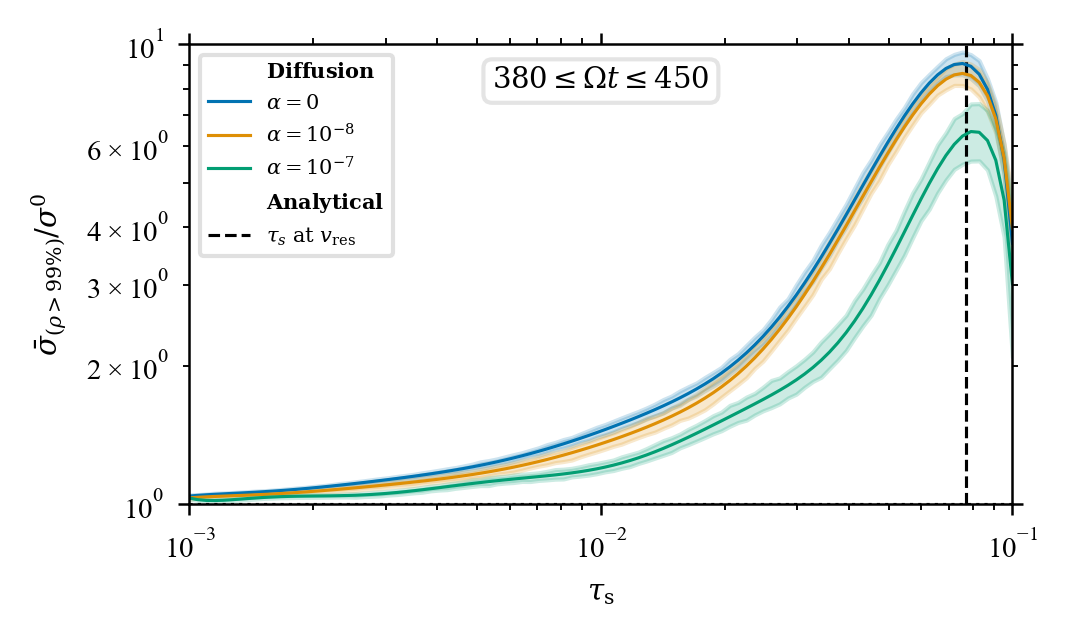}
    \caption{The normalized mean size distribution at the $99^\mathrm{th}$ percentile and between $380 \! \leq \! \Omega t \! \leq \! 450$ for a dust diffusion coefficient of $\alpha = 0 $ (run \texttt{PSI10}) in blue, $\alpha = 10^{-8} $ (run \texttt{PSI}$_{\alpha,1\mathrm{e}-8}$) in orange and $\alpha = 10^{-7} $ (run \texttt{PSI}$_{\alpha,1\mathrm{e}-7}$) in green.}
    \label{fig:diffusion_sizedist}
\end{figure}

The peak in the size distribution exists for all diffusion coefficients where the PSI was able to form clumps (Figure \ref{fig:diffusion_sizedist}). This means that the peak is independent of numerical or physical effects at the smallest scales, which are strongly damped by diffusion.

\subsection{Stokes range} \label{sec:stokesrange}

\begin{figure}
    \centering
    \includegraphics[width=1\linewidth]{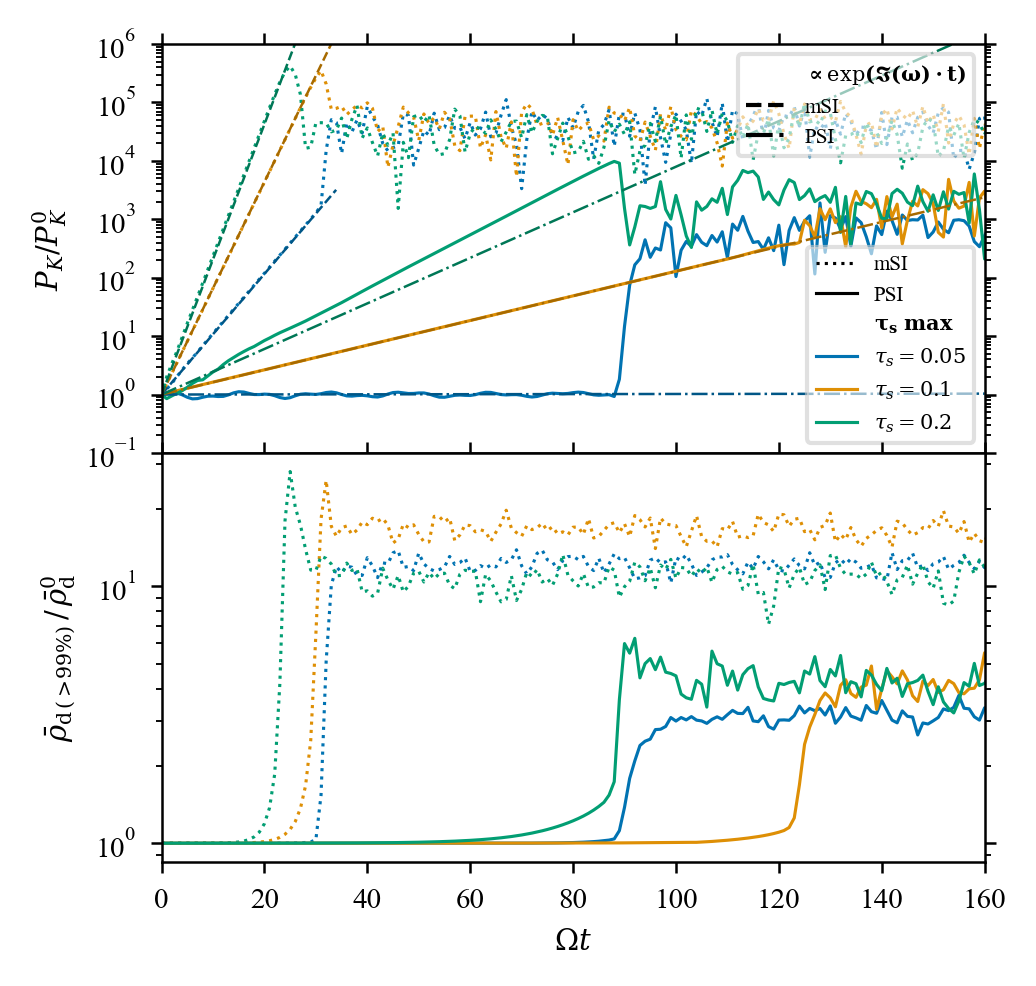}
    \caption{Time evolution of the mSI (runs: $\left[\texttt{mSI}^*, \texttt{mSI}_{\tau_\mathrm{s}, 5\mathrm{e}-2}, \texttt{mSI}_{\tau_\mathrm{s}, 2\mathrm{e}-1}\right]$) with a dotted line and PSI (runs: $\left[\texttt{PSI}^*, \texttt{PSI}_{\tau_\mathrm{s}, 5\mathrm{e}-2}, \texttt{PSI}_{\tau_\mathrm{s}, 2\mathrm{e}-1}\right]$) with solid lines for different peak Stokes numbers $\tau_\mathrm{s, \, max}$. Where the top plot shows the amplitude of the largest mode in the shearing box (A sinusoid with wavenumber $K_{x,z}$ = 30), and the dashed lines and dash-dotted lines showing the analytical results calculated with \texttt{psitools} for the corresponding $\alpha$. The bottom plot shows the normalized mean density at the $99^\text{th}$ percentile for every snapshot.}
    \label{fig:ts_peak_dens}
\end{figure}

The larger dust species have a bigger impact on the momentum transfer of the PSI. This means that the location of the discontinuous upper boundary of the size distribution will impact the PSI. For the mSI, the Stokes number influences the growth rate, lower Stokes numbers have lower growth rates, see top plot of Figure \ref{fig:ts_peak_dens}. This trend is also visible for the PSI, but the run \texttt{PSI}$_{\tau_\mathrm{s}, 5\mathrm{e}-2}$ did not grow at the perturbed wave of $K_{x,z}=30$, although around a time of $90 \, \Omega t$ there is growth. This could be explained by small numerical errors that excite a different wavenumber $\mathbf{K}$ than the initial perturbation that does have a higher growth rate. The saturated amplification factor is not strongly affected by the upper boundary of the size distribution (bottom plot Figure \ref{fig:ts_peak_dens}).

\begin{figure}
    \centering
    \includegraphics[width=1\linewidth]{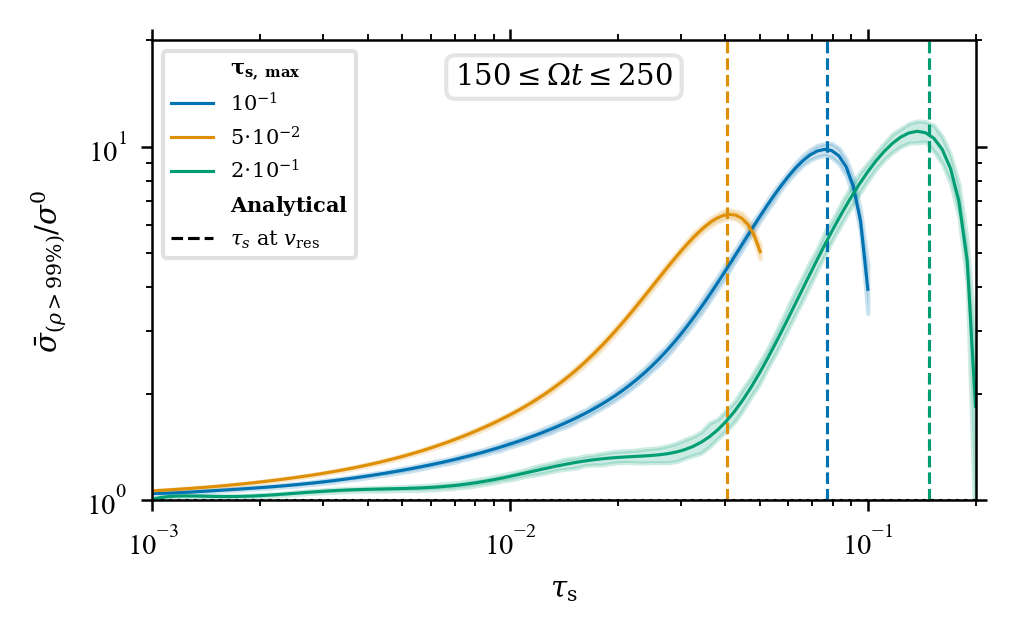}
    \caption{The normalized mean size distribution at the $99^\mathrm{th}$ percentile and between $380 \! \leq \! \Omega t \! \leq \! 450$ for a dust diffusion factor of $\alpha = 0 $ (run \texttt{PSI10}) in blue, $\alpha = 10^{-8} $ (run \texttt{PSI}$_{\alpha,1\mathrm{e}-8}$) in orange and $\alpha = 10^{-7} $ (run \texttt{PSI}$_{\alpha,1\mathrm{e}-7}$) in green.}
    \label{fig:ts_peak_sizedist}
\end{figure}

Changing the maximum Stokes number $\tau_{\mathrm{s, max}}$ will affect the size distribution in the upper $99^\text{th}$ percentile. The peak of the size resolution in standard run \texttt{PSI10} lies outside of the Stokes range of run \texttt{PSI}$_{\tau_\mathrm{s}, 5\mathrm{e}-2}$, but in this run, we still observe a peak in the distribution now at a different location. This means that the peak location is dependent on the initial size distribution and is always slightly smaller than the maximum Stokes Number and thus dependent on $\tau_{\mathrm{s, max}}$. Changing the Stokes range also changes the bulk properties of the distribution, like the average Stokes number $\bar{\tau}_\mathrm{s}$ and the average drift velocity $\bar{u}^0_x(\sigma)$. The bulk drift velocity corresponds to a resonance size $\tau_\mathrm{s}$ where ${v_\mathrm{d}}^0_x(\tau_\mathrm{s}) = \bar{u}^0_x(\sigma)$, see Section \ref{sec:evolution_sizedist}. This resonance size changes for the different Stokes ranges and is indicated with a vertical dashed line in Figure \ref{fig:ts_peak_sizedist}. The location of the peak could also depend on the resonant size. Distinguishing between the resonant size and maximum Stokes number can not be done by only changing the range of Size distribution, but we change the resonant size without changing the maximum Stokes number by changing the slope of the size distribution (Section \ref{sec:slope}). Another thing to take into account is that if the maximum Stokes number of the size distribution gets close to unity or even higher, the fluid approximation of the dust particles may not hold anymore.

\subsection{Slope of size distribution}\label{sec:slope}

\begin{figure}
    \centering
    \includegraphics[width=1\linewidth]{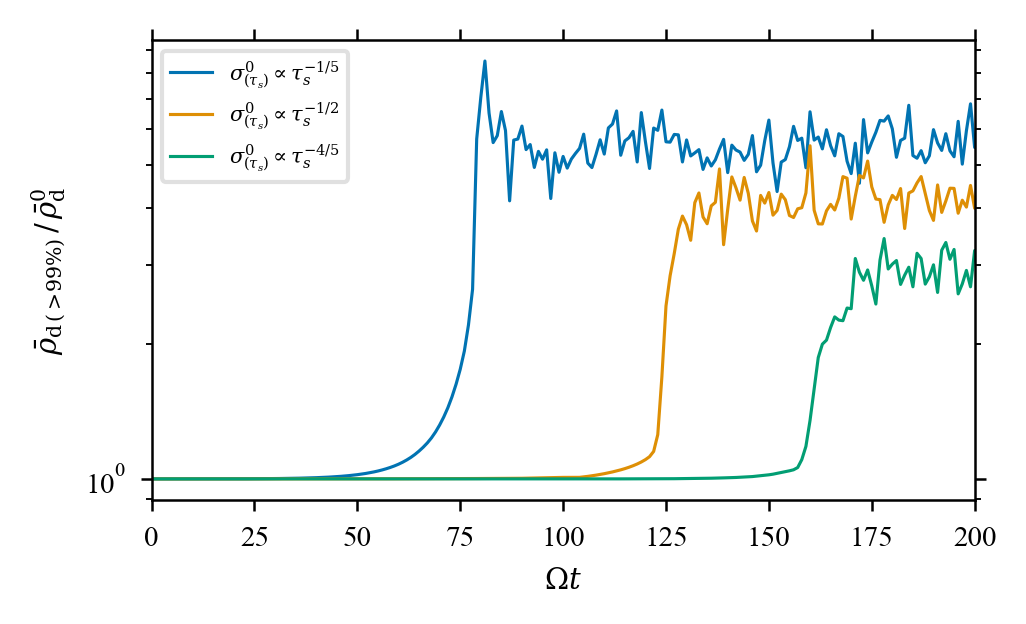}
    \caption{The normalized mean density at $99^\mathrm{th}$ percentile for different size distributions run \texttt{PSI}$_{\beta, -3.2}$ in blue ($\beta=-3.5$), run \texttt{PSI10} in orange and run \texttt{PSI}$_{\beta, -3.8}$ in green.}
    \label{fig:dens_slope}
\end{figure}

Varying the slope of the size distribution will change the bulk properties of the distribution without changing the Stokes range. Decreasing the slope of the size distributions power law will skew the distribution to have more mass and momentum transfer at a Stokes numbers closer to the upper boundary $\tau_\mathrm{s, \, max}$. This causes the PSI to be more similar to mSI and just like the mSI, the PSI with the smaller slope run \texttt{PSI}$_{\beta, -3.2}$ has a higher growth rate and saturated amplification factor than run  \texttt{PSI10} \& \texttt{PSI}$_{\beta, -3.8}$, see Figure \ref{fig:dens_slope}.
Similar to run  \texttt{PSI}$_{\tau_\mathrm{s}, 5\mathrm{e}-2}$, run \texttt{PSI}$_{\beta, -3.8}$ grows very slowly at $\mathbf{K}=\left(30,0,30\right)^T$ but get excited at another higher wavenumber by small numerical errors that overtakes the perturbed wave at $\mathbf{K}=\left(30,0,30\right)^T$.

\par
\begin{figure}
    \centering
    \includegraphics[width=1\linewidth]{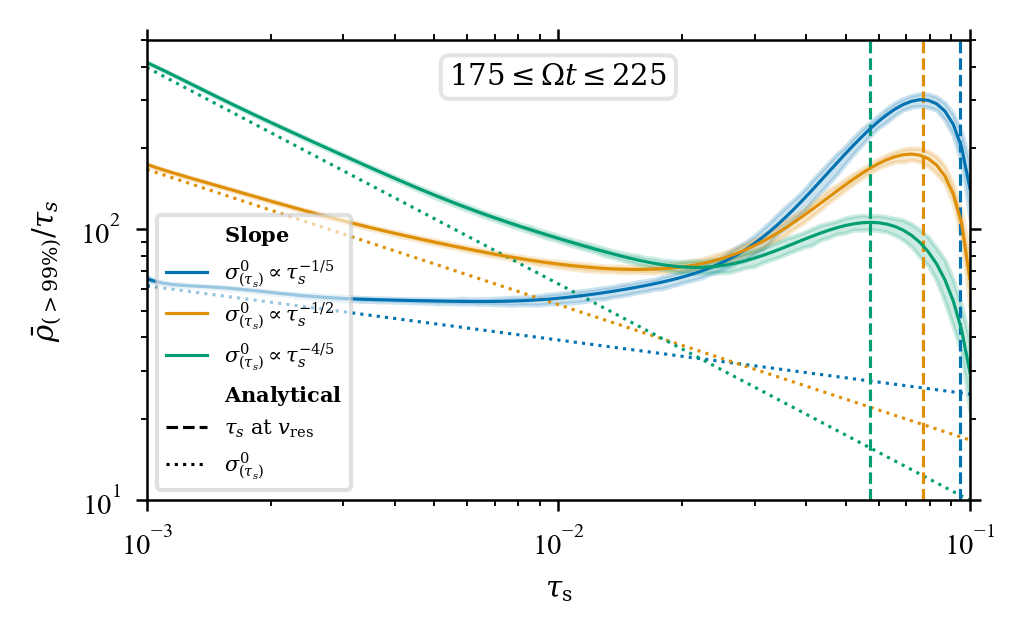}
    \caption{The mean size distribution at the $99^\mathrm{th}$ percentile and between $175 \! \leq \! \Omega t \! \leq \! 225$ for different power law slopes with $\beta = -3.2 $ (run \texttt{PSI}$_{\beta, -3.2}$ in blue, $\beta = -3.5$ (run \texttt{PSI10}) in orange and $\beta = -3.8$ (run \texttt{PSI}$_{\beta, -3.8}$) in green.}
    \label{fig:size_slope}
\end{figure}

 The resonant Stokes number corresponds to the resonant velocity of the whole size distribution dependent on the slope (${v_\mathrm{d}}^0_x(\tau_\mathrm{s}) = \bar{u}^0_x(\sigma)$), $\tau_\mathrm{s}$ at $v_\mathrm{res}$ is smaller for smaller $\beta$. If the location of the peak in the size distribution in upper percentiles is dependent on the resonant Stokes number, the peak would shift for different initial slopes of the size distribution. In Figure \ref{fig:size_slope}, we see that the location of the peak in the size distribution does not shift to the same extent as the Stokes number corresponding to the resonant velocity. The shift in the location of the peak is too small to quantify if the location of the peak is dependent on the resonant velocity. 

\subsection{Dust-to-gas ratio}\label{sec:dust_gas_ratio}

\begin{figure}
    \centering
    \includegraphics[width=1\linewidth]{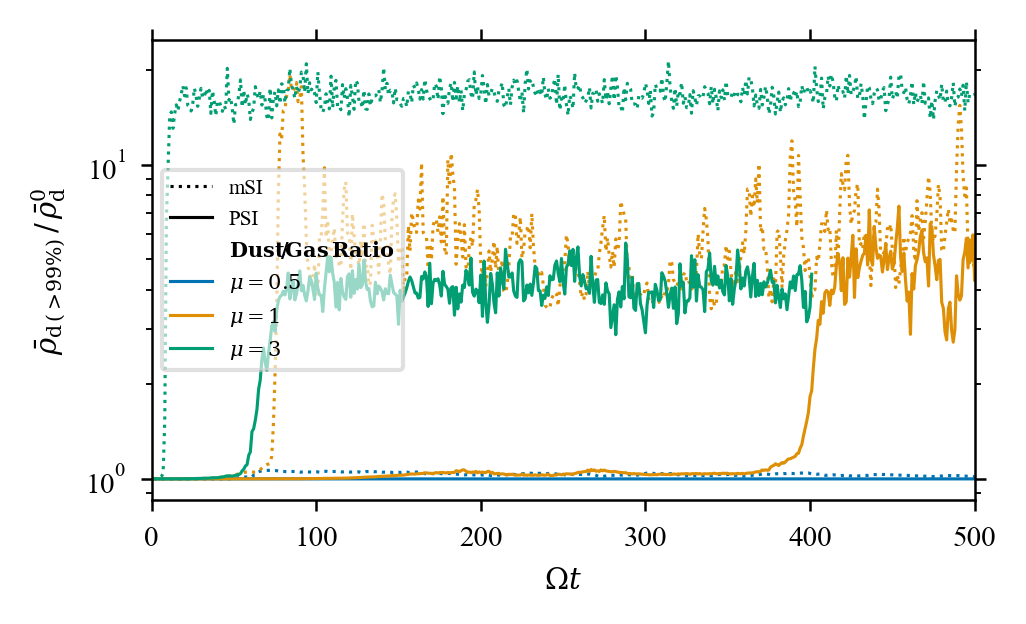}
    \caption{The normalized mean density at $99^\mathrm{th}$ percentile for different dust-to-gas ratios, with $\mu = 3$ (run \texttt{PSI}$_{\mu, 3}$ in green, $\mu = 1$ (run \texttt{PSI}$_{\mu, 1}$) in orange and $\mu = 0.5$ (run \texttt{PSI}$_{\mu, 0.5}$) in green.}
    \label{fig:dust_gas_ratio}
\end{figure}

The SI in the low dust-to-gas ratio regime $\mu < 1$ is a Resonant Drag Instability (RDI) \citep{Squire_a_2018ApJ...856L..15S}, and we can see from Figure A2 in \citet{McNally2021MNRAS.502.1469M} that there is no substantial growth for the PSI in the low dust-gas-ratio regime ($\mu=0.5$). This is a direct consequence of the adverse effect of the size distribution on the RDI streaming instability (Paardekooper \& Aly (in prep.)). We can compare this to numerical simulations at different dust-to-gas ratios that, if perturbed with white noise, should be able to grow in their fastest-growing mode. This is done with three different dust-gas ratios in runs \texttt{PSI}$^*_{\mu, 3}$, \texttt{PSI}$^*_{\mu, 1}$ and \texttt{PSI}$^*_{\mu,0.5}$, these simulations are perturbed by white noise in the gas, where the standard deviation is $10^{-4} \cdot c_\mathrm{s}$. When we perturb the gas with white noise, the run at $\mu=3$ reaches the saturated non-linear regime after $\Omega t \sim 75$, compared to $\Omega t \sim 125$ for the standard run (\texttt{PSI10}) where we only perturbed the wavevector $\mathbf{K}=\left(30,0,30\right)^T$. Similar to previous work \citep[e.g.][]{Yang2021MNRAS.508.5538Y, Zhu2021MNRAS.501..467Z, McNally2021MNRAS.502.1469M, Krapp2019ApJ...878L..30K}, lowering the dust-to-gas ratio will also decrease the growth rate. The run at a dust-to-gas ratio $\mu=1$ (\texttt{PSI}$^*_{\mu,1}$) takes more than four times as long to reach the saturated non-linear regime at $\Omega t \sim 400$. However, in contrast to the mSI case, the saturated amplification factor at the $99^\text{th}$ percentile has a similar saturated value at $\mu=1$ as the one of $\mu=3$, see Figure \ref{fig:dust_gas_ratio}. The normalized size distribution at the $99^\text{th}$ percentile of $\mu=1$ is also very similar to $\mu=3$, showing the same peaked structure. At a dust-to-gas ratio of $\mu=0.5$, there is no significant growth of the instability after the full length of the run $\Omega t = 500$.

\section{Substructure in clumps}\label{sec:substructure}

\begin{figure*}
    \centering
    \includegraphics[width=.9\hsize]{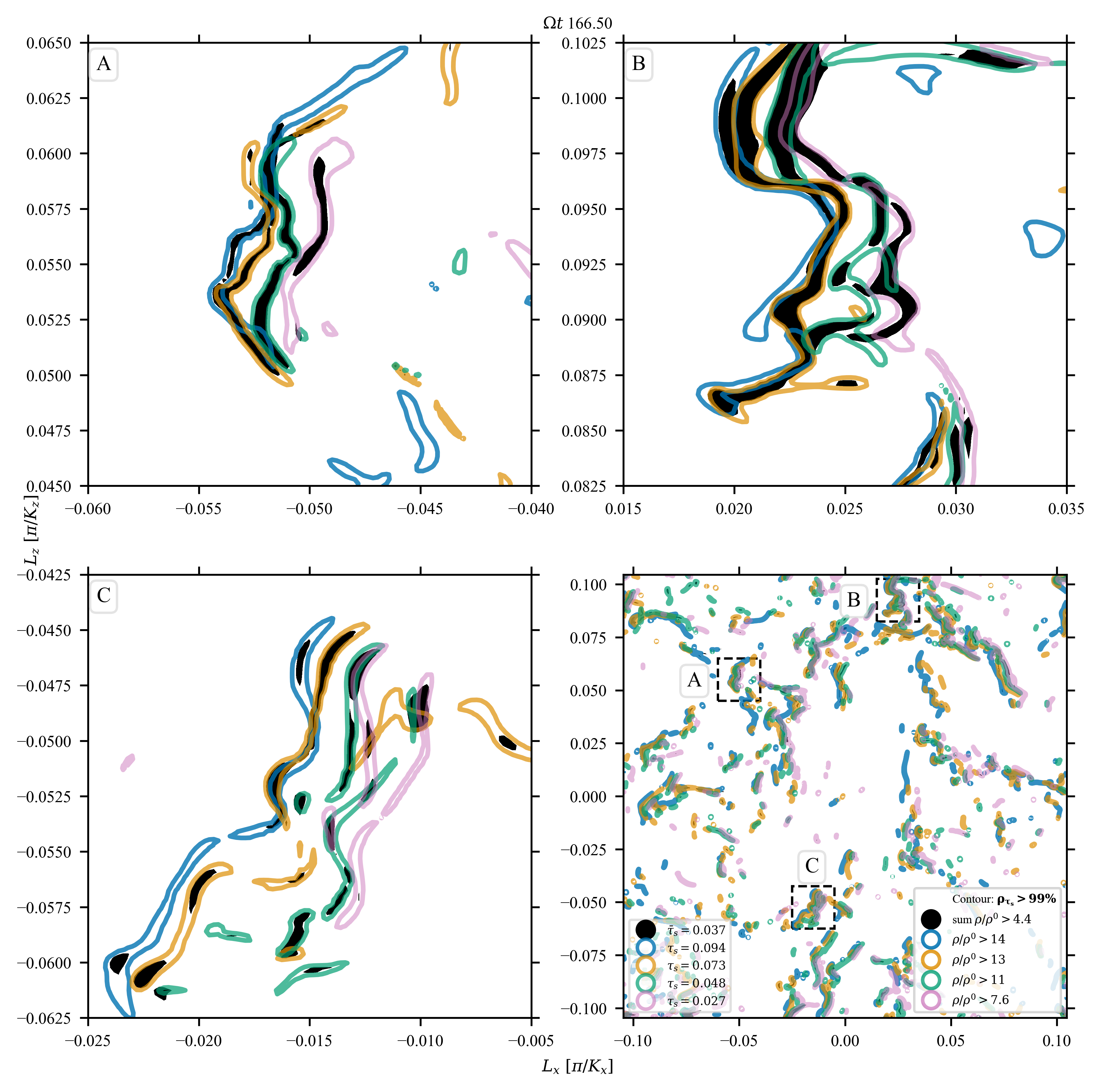}
    \caption{The contour of density at the upper $99^\text{th}$ percentile for the sum of the density and the dust bins of the four largest Stokes numbers. Showing the substructure of the clumps in the nonlinear regime (snapshot at $\Omega t=166.5$) for the PSI (run \texttt{PSI10}$_\texttt{1024}$).}
    \label{fig:substructure}
\end{figure*}

The peak in the size distribution at the upper $99^\text{th}$ percentile indicates that there is some substructure in clumps. Part of the substructure can be explained by the difference in the density distribution for the different Stokes numbers (Figure \ref{fig:poly10}D). Causing the general trend that the larger Stokes are overrepresented within the $99^\text{th}$ percentile of the sum of dust species (the area of $99^\text{th}$ percentile contour of the larger Stokes numbers is smaller than the sum of dust and area of the smaller Stokes numbers is larger). This only leads to the overabundance of larger dust sizes in the clumps but does not explain the peak in the size distribution. This is also visible in the dust-size distribution in Figure 9a of \citet{Yang2021MNRAS.508.5538Y}, where high-$\mu$ run (Af) shows an over-representation of the largest dust sizes in the denser regions.

\par An effect that can lead to the substructure and a peak in the size distribution is a spatial separation between the highest-density regions of different Stokes. This Stokes-dependent separation can occur through the PSI, where the different dust sizes undergo different drag forces when drifting through the gas. Causing the most decoupled dust sizes (largest Stokes numbers) to be in front of the smaller Stokes sizes in the clumps in the drift direction. We can visualize this by plotting the contour of the densities $99^\text{th}$ percentile for the different dust species. This is done for the four largest dust sizes and the sum over all dust species in Figure \ref{fig:substructure}. In this figure, we can see that the individual clumps are made of adjacent filaments of the individual dust species, where we find these filaments are in order of size, with often the largest Stokes number on the left. This corresponds to the direction of the background pressure gradient and drift direction.

\par The density contour indicates where most of a specific dust is located, but the density and size distribution are more discontinuous. In the numerical simulation, we always need to use a discrete number of dust species ($n_\mathrm{d} = 10$ for run \texttt{PSI10}$_\texttt{1024}$). Therefore, we see some gaps between the contours. These become smaller if we take the limit to a continuous size distribution ($n_\mathrm{d} \rightarrow \infty$), but the averaged behaviour of the clumps is already converged at $n_\mathrm{d} = 10$ (Section \ref{sec:ndustspecies}). Within a contour of a dust bin, the size distribution is not monodisperse. However, the spatial dispersion in the clumps causes specific regions to be more monodisperse compared to the background distribution.

\par The spatial separation also causes the peak size distribution in the highest percentiles because the largest Stokes numbers are slightly in front of the rest of the clump, causing part of the dust to fall outside of the percentile contour of the sum of dust species. More of the largest dust species are slightly ahead of the rest of the clump and, therefore, less present in the densest regions of the sum of the dust species. This showed up in the averaged size distribution at $99^\text{th}$ percentile in density as a peak (see Figure \ref{fig:sizedistribution}).

\par When we take the average size distribution over the contour of individual clumps, we find that the individual clumps also form a peak, and the shape of the size distribution is consistent with the shape of the average size distribution, indicating that the peak is not a symptom of averaging over high-density regions that have different individual compositions.

\section{Discussion and Conclusion}\label{sec:discussion}

\subsection{Maximum dust density}

\par One of the main results of this paper is that the maximum density of the dust is significantly lower for the PSI than the mSI $\sim \mathcal{O}(10)$; this will make it harder to reach a large enough amplification, where clumping can lead to planetesimal formation. The natural clumping criteria in 3D is the Roche density,
\begin{align}
    \rho_\mathrm{R} = \frac{9 \Omega^2}{4 \pi G},
\end{align}
with clumping expected for $\rho>\rho_\mathrm{R}$. We follow \cite{Li2021ApJ...919..107L, Lim_b2024arXiv241017319L} and apply the same criteria in 2D, but it should be noted that 3D simulations are required to assess clumping using the Roche density. When assuming a low mass disc with a Toomre parameter $Q$ of 32 \citep{Toomre1964ApJ...139.1217T}, the Roche density $\rho_\mathrm{R} \approx 180 \rho^0_\mathrm{g}$. The mSI runs reach the Roche density, but in the PSI simulations we reach a maximum density of $\rho_\mathrm{d, max} = 82 \pm 14 \rho^0_\mathrm{g}$ (see run \texttt{PSI10}$_{1024}$ in Figure \ref{fig:poly10}), which would be below the strong clumping criteria of \citet{Li2021ApJ...919..107L,  Lim_b2024arXiv241017319L}, making it more difficult to form planetesimals through the PSI compared to the mSI. However, since the dust is modelled as a pressureless fluid, a higher spatial resolution will lead to larger densities, possibly exceeding the Roche density.

\par Another important consideration is that the model we used is simplified in several ways. First of all, we considered only two spatial dimensions. Therefore, the highest-density regions are not true clumps in 3D but are sheets of higher density in 2D. To represent real clumps that could end up becoming planetesimals, we need to consider the 3D simulations; this could affect the shape of the clumps and, in turn, also affect the spatial separation we observe in the highest-density regions between the different dust species. 

\par Secondly, the simulations do not consider self-gravity. This does not prevent us from comparing to clumping criteria; see, for example, \citet{Li2021ApJ...919..107L}, they studied clumping in 2D simulations without doing self-gravity using the Roche density as criteria. \citet{Lim_a2024ApJ...969..130L}\footnote{Section 3.2} also shows that switching on self-gravity in 3D simulations of the SI only significantly affects density structure if the maximum density already exceeds the Roche density. Although self-gravity could affect the substructure in the formed clumps, studying the PSI without self-gravity would not significantly change the clumping criteria. 

\par The biggest simplification is that our simulations are \emph{unstratified}, while previous simulations of the multi-species streaming instability have mostly been \emph{stratified} 3D simulations \citep{Bai2010ApJ...722.1437B, Schaffer2018A&A...618A..75S, Schaffer2021A&A...653A..14S, Rucska2023MNRAS.526.1757R}. Although these simulations show similar trends, like larger dust species playing a larger role in the formed structure, they can not directly be compared with the unstratified simulations in this paper or from \citet{Yang2021MNRAS.508.5538Y}. \citet{Lin2021ApJ...907...64L} shows that the dominant instability in stratified discs is driven by the vertical gradient in the azimuthal velocity of the dusty gas (vertical shear), which provides a source of "free energy" resulting in an instability through partial dust-gas coupling. The vertically shearing streaming instability (VSSI) has fast growth rates compared to the "classical" SI and will likely form the initial turbulence in stratified simulations. \citet{Li2021ApJ...919..107L} find that the linear unstratified SI growth rates are not a good predictor for SI clumping in stratified simulations and that analytical growth rates are surprisingly similar to non-clumping runs.

\par Lastly, the simulations in this paper are isothermal; according to \citet{Lehmann2023MNRAS.522.5892L}, the non-isothermal SI will suppress sufficiently small-scale modes through radial buoyancy only if the gas cooling timescale is comparable to the dynamical one ($\beta \approx 1$). However, the linear analysis in this paper has been done at larger scale modes ($K=30$) and is less affected if we assume the SI occurs in specific regions where the cooling occurs on timescales roughly similar to dynamical. Whether this affects the clumping would have to be tested with non-isothermal simulations.

\subsection{Gauss-Legendre integration}

\par We have introduced a new way to implement the integration of the backreaction on the gas, where we take the continuum limit using the GL method to sample the size distribution. We can compare the error in the momentum transfer from the integration \eqref{eq:interal_equilibrium_solution} in the linear regime, where we found that the GL method converges a lot faster to the continuum limit than the discrete method. The GL method is shown to work well in the linear regime (see Figure \ref{fig:sample_sizedistribution}) where the size distribution is a smooth function, and the momentum transfer integral is well-behaved and thus can be accurately approximated by the GL nodes. Even though this expectation is no longer valid in the non-linear regime, we notice that the size distribution resulting from the GL integration remains smooth (see Figure \ref{fig:sizedistribution}) and converges with the number of dust species used in the sampling (see Figure \ref{fig:ndust_sizedist}). 

An important distinction is that we take a truly polydisperse approach, i.e., with a continuum in sizes, using the GL method. There have also been multi-species SI studies using different numerical integration routines. \citet{Bai2010ApJ...722.1437B} use a particle-fluid approach where the continuous size distribution is discretized into a number of bins with a fixed width in Stokes number in the logarithmic scale (each bin covers half a $\mathcal{O}(10)$), and where equal mass bins are also assumed. A similar method (discrete logarithmic sampling) is used in \citet{Krapp2019ApJ...878L..30K, Schaffer2018A&A...618A..75S, Zhu2021MNRAS.501..467Z, Yang2021MNRAS.508.5538Y}. \citet{Rucska2023MNRAS.526.1757R} also uses equal mass boundaries, but these are not equally spaced linearly or logarithmically. This discrete sampling method is analogous to using a Riemann sum approximation with a midpoint rule and is also bound by the approximation error of this routine. Therefore, at a low number of dust species, this discrete method does not represent the continuous limit, and the integration error between the discrete method and the continuous limit converges significantly slower with the number of dust species than the GL method (see section \ref{sec:linear}).

Simulating the PSI using a particle-fluid approach, the dust size distribution can also be quasi-continuously, where each particle is given a unique size sampled from the size distribution. This method is explored in \citet{Schaffer2018A&A...618A..75S, Schaffer2021A&A...653A..14S}, and is analogous to a Monte-Carlo algorithm for the numerical integration of the momentum transfer, which converges even slower than the midpoint rule.

\subsection{Effect on planetesimals}

\par Our analysis revealed that in the densest regions, larger dust sizes are overrepresented, and there is a peak in the size distribution. This peak was not expected, and we explored a couple of explanations, like the resonant velocity of the size distribution and the spatial separation of the different dust species. The spatial separation is the preferred explanation for the peak in the densest regions, causing the largest dust species to fall outside of the densest region of the sum over all the dust. The observed size distribution in these densest regions can mimic the distribution that can result from coagulation/dust growth \citep{Birnstiel2011A&A...525A..11B}. We can form the peak in the sized distribution through dynamics and the size-dependent coupling between the gas and the dust, without dust growth or fragmentation. That the size distribution changes through dynamics can in turn also affect coagulation, by e.g. making the size distribution locally more monodisperse. Therefore, combining coagulation with polydisperse hydrodynamical simulations of the SI would be an important further step.

\par If the peak in the size distribution persists when the density exceeds the Roche density, the peak in the size distribution also ends up in rubble pile asteroids \citep{Chapman1978NASCP2053..145C, Walsh2018ARA&A..56..593W, Visser2021A&A...647A.126V}. The size distribution of the particles making up the rubble pile can be observed \citep[e.g.][]{Gundlach2013Icar..223..479G, Blum2017MNRAS.469S.755B, Fulle2017MNRAS.469S..39F}. The observed size distribution can then inturn be an indication of the formation history and the local environment where the asteroid was formed through gravitational collapse, taking into account how the PSI changes the size distribution in the highest-density regions.  However, we then need to consider that the size of the aggregates can change during the collapse \citep{Wahlberg2017MNRAS.469S.149W, Pinto2021ApJ...917L..25P, Visser2021A&A...647A.126V}, although the size distribution at the Roche density can also affect the cloud collapse. The size distribution can also change at later stages due to radiative or chemical reactions or kinetic impacts \citep{Poulet2016MNRAS.462S..23P, Graves2019Icar..322....1G, Hsu2022NatAs...6.1043H}. Although this primarily affects the surface, the interior of the asteroid, below a layer of a few tens of centimetres, can still represent the primordial size distribution \citep{Capria2017MNRAS.469S.685C}.

\subsection{Summary}

\par The main results and findings of this paper are:
\begin{itemize}
    \item The Gauss-Legendre quadrature method used for sampling the size distribution converges much faster with the number of dust species compared to a discrete method when we compare the growth rates of the numerical simulations to the analytical growth rate (see section \ref{sec:linear}). This allows us to use fewer dust species whilst maintaining an acceptable accuracy on the momentum transfer between the dust and the gas in the continuum limit, which can be significantly less computationally expensive.
    \item Similar to previous work \citep{Krapp2019ApJ...878L..30K, Paardekooper2020MNRAS.499.4223P, Paardekooper2021MNRAS.502.1579P, McNally2021MNRAS.502.1469M, Zhu2021MNRAS.501..467Z}, we find that the PSI has slower growth rates than mSI. This makes the parameter space where the PSI can be triggered to form clumps smaller, especially when we consider that the growth rate of PSI changes faster with maximum Stokes number or dust diffusion than the mSI (see section \ref{sec:linear} and \ref{sec:diffusion} for more details).
    \item In the regimes where the PSI forms clumps, the amplification factor can be up to an order of ten smaller than the mSI at the maximum Stokes number of PSI range, making it more difficult for the clumps to reach the Roche density limit and collapse into planetesimals.
    \item The PSI can \emph{not} be represented by an mSI run with properties that are derived from the averaged properties of the PSI setup such as average Stokes number and total dust density of the size distribution, and the PSI is not a simple summation of individual discrete mSI cases at different Stokes numbers.
    \item There is a correlation between the location of higher density regions of the different dust sizes in a PSI run, although the larger dust sizes clump more. There is a pronounced overrepresentation of larger dust species in the dense regions in the non-linear regime. This overrepresentation occurs because smaller dust particles are more tightly coupled to the nearly incompressible gas, which limits their ability to clump as effectively as the larger dust species.
    \item We observe a peak in the size distribution of the densest regions. This peak arises from the spatial segregation of the differently sized dust species, where particles with the largest Stokes numbers are located just outside the densest areas of the total dust density (the combined dust species). This spatial separation plays a crucial role in shaping the size distribution in these regions.
    \item The peak in the size distribution at the densest regions can mimic the distribution we see when we have coagulation/dust growth \citep{Birnstiel2011A&A...525A..11B}. We find that through dynamics and size-dependent coupling between the gas and dust, we can form a size distribution with a bump on top of the background MRN distribution without dust growth or fragmentation.
\end{itemize}

Data Availability: The data that support the findings of this study are openly available in 4TU.ResearchData under the name: Data underlying the manuscript "Polydisperse Formation of Planetesimals" at \url{http://doi.org/10.4121/8f0a99b2-7d55-4e05-aa8f-345cab38ff38}. The code that processes the output of \texttt{FARGO3D} and supports the findings of this study is openly available in 4TU.ResearchData under the name: Code underlying the manuscript "Polydisperse Formation of Planetesimals" at \url{http://doi.org/10.4121/528086f3-b288-482f-ab45-3c5ccb18a293}.

\begin{acknowledgements}
We would like to thank the anonymous referee for their invaluable feedback. We would like to thank Benjamin Silk and Tom Konijn for their useful comments. This project has received funding from the European Research Council (ERC) under the European Union’s Horizon Europe research and innovation programme (Grant Agreement No. 101054502). The authors acknowledge the use of computational resources of the DelftBlue supercomputer, provided by \citet{DHPC2024}. This work made use of several open-source software packages. We acknowledge \texttt{numpy} \citep{Harris2020Natur.585..357H}, \texttt{matplotlib} \citep{Hunter2007CSE.....9...90H}, \texttt{scipy} \citep{Virtanen2020NatMe..17..261V}, \texttt{FARGO3D} \citep{Benitez-Llambay2016ApJS..223...11B, Benitez-Llambay2019ApJS..241...25B}, and \texttt{psitools} \citep{Paardekooper2020MNRAS.499.4223P, Paardekooper2021MNRAS.502.1579P, McNally2021MNRAS.502.1469M}. We acknowledge 4TU.ResearchData for supporting open access to research data.
\end{acknowledgements}

%
%

\bibliographystyle{aa_url.bst} 
\bibliography{aanda.bib} 

\end{document}